\documentclass[11pt,preprint, preprintnumbers,showpacs, showkeys, aps, prd, amsmath, amsfonts, amssymb,
   eqsecnum, nofootinbib, floatfix, secnumarabic]{revtex4}
   
\usepackage{epsf}
\usepackage{epsfig}
\usepackage{graphics}
\usepackage{graphicx}
\usepackage{amssymb}

\def\lsim{\mathrel{\raise.3ex\hbox{$<$\kern-.75em\lower1ex\hbox{$\sim$}}}}
\def\gsim{\mathrel{\raise.3ex\hbox{$>$\kern-.75em\lower1ex\hbox{$\sim$}}}}

\begin{document}
\preprint{CUMQ/HEP 180}

\title{ \Large  Vector Leptons in the Higgs Triplet Model }

\author{Sahar Bahrami$^1$\footnote{Email: sahar.bahrami@concordia.ca}}
\author{Mariana Frank$^1$\footnote{Email: mariana.frank@concordia.ca}}

\affiliation{ $^1 $Department of Physics,  
Concordia University, 7141 Sherbrooke St. West ,
Montreal, Quebec, Canada H4B 1R6.}

\date{\today}

\begin{abstract}

We analyze the phenomenological implications of introducing vector-like leptons on  the Higgs sector  in the Higgs Triplet Model. We impose only a parity symmetry which disallows mixing between the new states and the ordinary leptons. If the vector leptons are allowed to be relatively light, they enhance or suppress   the decay rates of loop-dominated  neutral Higgs bosons decays $h \to \gamma \gamma$ and $h \to Z \gamma$, and affect their correlation.  An important consequence is that, for light vector leptons, the decay patterns of the  the doubly-charged Higgs boson will be altered, modifying  the restrictions on their masses.

\pacs{14.80.Fd, 12.60.Fr, 14.60.Pq.}
\keywords{LHC phenomenology, Higgs Triplet Model}
\end{abstract}
\maketitle

 \section{Introduction}
 \label{sec:intro}
 With the discovery of the Standard Model (SM) Higgs-like scalar at the LHC \cite{HiggsLHC}, the SM particle content seems complete. In particular, the mass and couplings of the neutral Higgs boson seem to disfavor an additional chiral generation of quarks and leptons \cite{Aaltonen:2009nr}. However, additional vector-like fermions, in which an SM generation is paired with another one of opposite chirality, and with identical couplings,  are less constrained, as there is no  quadratic contribution to their masses. These states appear as natural extensions of the SM particle content, in theories with warped or universal extra dimensions, as Kaluza-Klein excitations of the bulk field \cite{Agashe:2004cp}, in non-minimal  supersymmetric extensions of the SM  \cite{Moroi:1991mg}, in composite Higgs models \cite{Dobrescu:1997nm}, in Little Higgs Model \cite{ArkaniHamed:2002qy} and in gauged flavor groups \cite{Davidson:1987tr}. Vector fermions have identical left- and right-handed couplings and can have masses which are not related to their couplings to the Higgs bosons \cite{Azatov:2009na}. Depending on the dominant decay mode, the limits on new vector-like fermions range from $\sim 100-600$ GeV \cite{VFlim}, rendering them observable at the LHC.
 
 Vector quarks can modify both the production and decays of the Higgs boson at the LHC, while vector leptons do not carry $SU(3)_c$ charge and can only modify the decay patterns of the Higgs. Study of the latter would be a sensitive probe for new physics. The lepton states contribute to self-energy diagrams for electroweak gauge boson masses and precision observables, and consistent limits on their masses and mixings have been obtained \cite{delAguila:2008pw,Joglekar:2012vc}.   
  
Vector  leptons have been studied in the context of the SM \cite{Joglekar:2012vc,Kearney:2012zi,Ishiwata:2011hr}, but less so for models beyond the SM, where they can also significantly alter the phenomenology of the model.  In the SM, introducing heavy fermions provides a contribution of the same magnitude and sign to the one of the top quark and interferes destructively with the dominant  $W$-contribution,  reducing the  $h\to \gamma \gamma$ rate with respect to its SM value. Recent studies indicate that cancellations between scalar and fermionic contributions allow a wide range of Yukawa and mass mixings among vector states \cite{Garg:2013rba}.   
An investigation of vector leptons in the two Higgs doublet model \cite{Garg:2013rba} showed that the presence of additional Higgs bosons alleviates electroweak precision constraints. Introducing vector leptons into supersymmetry \cite{Joglekar:2013zya} can improve vacuum stability and enhance the di-photon rate by as much as 50\%, while keeping new particle masses above 100 GeV and preserving vacuum stability conditions.

In the present work, we investigate vector leptons in the context of the Higgs Triplet Model (HTM). We do not deal with LHC phenomenology (pair production and decay) of the extra leptons, which has been discussed extensively in the literature \cite{Giudice:2008uua}, choosing instead to focus on signature features of this model. We have previously shown that in the Higgs Triplet Model, enhancement of the $h \to \gamma \gamma $ rate is possible only for the case where the doublet and triplet neutral Higgs fields mix considerably \cite{Arbabifar:2012bd}. We extend our analysis to include additional  vector-like leptons in the model and investigate how these affect  the Higgs di-photon decay rate, with or without significant mixing in the neutral Higgs sector. Originally, both CMS and ATLAS experiments at LHC observed an enhancement of the Higgs di-photon rate, while the di-boson rates ($h \to WW^*,\, ZZ^*$) have been roughly consistent with SM expectations. At present CMS  observes $\sigma(pp \to h) \times BR (h \to \gamma \gamma)=0.77 \pm 0.27$ times the SM rate, while ATLAS observes  $\sigma(pp \to h) \times BR (h \to \gamma \gamma)=1.55  ^{+0.33}_{-0.28}$ times the SM rate \cite{ATLASnew}. Given these numbers, it is possible that either the SM value will be proven correct, or a modest enhancement will persist.  A further test of the SM is the correlation of  the decay $h \to Z \gamma$ with one  for $h \to \gamma\gamma$. We also include the prediction for the vector lepton effect on  branching ratio of $h \to Z \gamma$ and comment on the relationship with the di-photon decay.

We have an additional reason to investigate the effects of vector leptons in the Higgs Triplet Model. The model includes doubly-charged Higgs bosons, predicted by most models to be light. Being pair-produced, the doubly-charged Higgs bosons  are assumed to
decay into a pair of leptons with the same electric charge, 
through Majorana-type interactions \cite{Chatrchyan:2012ya}.
Assuming a branching fraction of 100\% decays into leptons, {\it i.e.},
neglecting possible decays into  $W$-boson pairs,
the doubly-charged Higgs mass has been constrained to be larger than about 450 GeV, or more, depending on the decay channel. However, if the vector leptons are light enough, which  they can be,  the doubly-charged Higgs bosons can decay into them and thus evade the present collider bounds on their masses. We investigate this possibility in the second part of this work.

Our work is organized as follows. We introduce the Higgs Triplet Model with vector leptons in Section \ref{sec:model}. In Section \ref{sec:neutral} we analyze the effect of the vector leptons on the decays of the neutral Higgs bosons, and discuss the constraints on the parameter space coming from requiring agreement with present LHC data. We include both loop-dominated decays, $h \to \gamma \gamma$ in \ref{subsec:gamgam},  and $h \to Z \gamma $ in \ref{subsec:Zgam}. In Section \ref{sec:doublycharged} we analyze the effect of the vector leptons on the production and decay mechanisms of the doubly-charged Higgs at the LHC. We summarize our findings and conclude in Section \ref{sec:conclusion}.

 %
\section{The Model}
\label{sec:model}
The Higgs Triplet Model has been studied previously \cite{Akeroyd:2007zv}. Here we concentrate on the effect of extending the model by incorporating additional vector leptons. For the purpose of our analysis, vector quarks either do not exist, or are much heavier and decouple from the spectrum. The model contains a vector-like fourth generation of leptons, namely the $SU(2)_L$ left-handed lepton doublets $L_L^\prime=(\nu^\prime_L, e^\prime_L)$, right-handed charged and neutral lepton singlets, $\nu_R^\prime$ and $e_R^\prime$,  and the mirror right-handed lepton doublets, $L_R^{\prime\prime}=(\nu^{\prime\prime}_R, e^{\prime\prime}_R)$ and left-handed charged and neutral lepton singlets $\nu_L^{\prime\prime}$ and $e_L^{\prime\prime}$. The vector-like leptons have the following quantum numbers under $SU(3)_C \times SU(2)_L \times U(1)_Y$: 
\begin{eqnarray}
L^\prime_L=(\mathbf{1}, \mathbf{2}, -1/2), \qquad L^{\prime \prime}_R= (\mathbf{1}, \mathbf{2}, -1/2), \qquad e^\prime_R&=&(\mathbf{1}, \mathbf{1}, -1),\nonumber\\
 e^{\prime \prime}_L= (\mathbf{1}, \mathbf{1}, -1), \qquad  \qquad \nu^\prime_R=(\mathbf{1}, \mathbf{1}, 0), \qquad \nu^{\prime \prime}_L&=&(\mathbf{1}, \mathbf{1}, 0), 
\end{eqnarray}
with the electric charge given by $Q=T_3+Y$, where $T_3$ the weak isospin. The Lagrangian density of this model is: 
\begin{eqnarray}
\mathcal{L}_{\rm{HTM}}=\mathcal{L}_{\rm{kin}}+\mathcal{L}_{Y}+\mathcal{L}_{\rm VL}-V(\Phi,\Delta), 
\end{eqnarray}
where $\mathcal{L}_{\rm{kin}}$, $\mathcal{L}_{Y}$, $\mathcal{L}_{\rm VL}$ and $V(\Phi,\Delta)$ are 
the kinetic term, Yukawa interaction for the ordinary SM fermions, the mass and Yukawa interaction for the the vector leptons, and the scalar potential, respectively. 
 The Yukawa interactions for the ordinary SM leptons  are \cite{Arbabifar:2012bd} 
\begin{eqnarray}
\mathcal{L}_Y&=&-\left[\bar{L}_L^ih_e^{ij}\Phi e_R^j+\rm{h.c.}\right] -\left[ h_{ij}\overline{L_L^{ic}}i\tau_2\Delta L_L^j
+\rm{h.c.} \right],~~~~ \label{nu_yukawa}
\end{eqnarray}
where $\tilde{\Phi}=i\tau_2 \Phi^*$,  $h_{e}$ is a  3$\times$3 complex matrix, and $h_{ij}$ is a $3\times 3$ complex symmetric Yukawa matrix. 
Additionally, with the vector-like family of leptons as defined above, the vector-lepton part of the Lagrangian density is
\begin{eqnarray}
\mathcal{L}_{\rm VL}&=&-\big [M_{L}{\bar L}_L^\prime L_R^{\prime \prime}+M_{E}{\bar e}_R^{\prime} e_L^{\prime \prime}+ M_{\nu}{\bar \nu}_R^{ \prime} \nu_L^{\prime \prime}+\frac12 M_\nu^\prime {\overline \nu^{\prime c}_R} \nu_R^\prime + \frac12 M_\nu^{\prime \prime} {\overline \nu^{\prime \prime c}_L} \nu_L^{\prime \prime}+h_E^\prime ({\bar L}_L^\prime \Phi )e_R^\prime
 \nonumber \\
&& +h_E^{\prime \prime} ({\bar L}_R^{\prime \prime} \Phi )e_L^{\prime \prime}+h_\nu^\prime ({\bar L}_L^\prime \tau \Phi^\dagger )\nu_R^\prime +h_\nu^{\prime \prime} ({\bar L}_R^{\prime \prime} \tau \Phi^\dagger )\nu_L^{\prime \prime} + h^\prime_{ij}\overline{L_L^{\prime\, c}}i\tau_2\Delta L_L^{\prime\, }
+h^{\prime \prime}_{ij}\overline{L_R^{\prime \prime \, c}}i\tau_2\Delta L_R^{\prime\prime } \nonumber \\
&&+  \lambda_E^i ({\bar L}_L^\prime \Phi )e_R^i +  \lambda_L^i ({\bar L}_L^i \Phi )e_R^\prime + \lambda^\prime_{ij}\overline{L_L^{ ic}}i\tau_2\Delta L_L^{\prime}
+ \lambda^{\prime \prime}_{ij}\overline{L_R^{ ic}}i\tau_2\Delta L_R^{\prime \prime}   +\rm{h.c.} \big]
\end{eqnarray}
where we include explicit mass terms, Yukawa interactions among vector leptons, and Yukawa interactions between vector leptons and ordinary leptons.  The scalar potential is
\begin{eqnarray}
V(\Phi,\Delta)&=&m^2\Phi^\dagger\Phi+M^2\rm{Tr}(\Delta^\dagger\Delta)+\left[\mu \Phi^Ti\tau_2\Delta^\dagger \Phi+\rm{h.c.}\right]+\lambda_1(\Phi^\dagger\Phi)^2 \nonumber\\
&+&\lambda_2\left[\rm{Tr}(\Delta^\dagger\Delta)\right]^2 +\lambda_3\rm{Tr}[(\Delta^\dagger\Delta)^2]
+\lambda_4(\Phi^\dagger\Phi)\rm{Tr}(\Delta^\dagger\Delta)+\lambda_5\Phi^\dagger\Delta\Delta^\dagger\Phi,~~~~ \label{pot_htm}
\end{eqnarray}
with $m$ and $M$ the Higgs bare masses, $\mu$  the lepton-number violating  parameter,    and 
$\lambda_1$-$\lambda_5$ the Higgs coupling constants. The expressions for the $\lambda_1$-$\lambda_5$ parameters in terms of Higgs masses are given in \cite{Arbabifar:2012bd}.

The electroweak gauge symmetry is broken by the VEVs of the neutral components of the  doublet and triplet Higgs fields, 
\begin{eqnarray}
\langle \Phi^0 \rangle =v_\Phi/\sqrt{2}, \qquad \langle \Delta^0 \rangle= v_\Delta/\sqrt{2}.
\end{eqnarray}
where $\Phi$ and $\Delta$ 
are  the doublet Higgs field and the triplet Higgs field, with 
$v^2\equiv v_\Phi^2+2v_\Delta^2 =$ (246 GeV)$^2$. Higgs masses and coupling constants in the presence of non-trivial mixing in the neutral sector have been obtained previously  \cite{Arbabifar:2012bd}.

One can invoke new symmetries to restrict the interaction of the vector leptons with each other, or with the ordinary leptons, or disallow the presence of bare mass terms in the Lagrangian. For instance,
\begin{enumerate}
\item If  an additional $U(1)$ symmetry under which the primed,  double primed and the ordinary leptons have different charges, this would forbid explicit masses $M_L$, $M_E$, $M_\nu$ and $M^\prime_\nu$ in the Lagrangian. Vector leptons would get masses only through couplings to the Higgs doublet fields \cite{Ishiwata:2011hr,Dermisek:2013gta}. 
\item If one imposes a symmetry under which all the new $SU(2)$ singlet fields are odd, while the new $SU(2)$ doublets are even, this forces all  Yukawa couplings involving new leptons to vanish,  $h_E^\prime =h_E^{\prime \prime}=h_\nu^\prime=h_\nu^{\prime \prime}=h_{ij}^{\prime}=h_{ij}^{\prime \prime}=0$, and the masses arise only from explicit terms in the Lagrangian \cite{Joglekar:2012vc}.
\item Finally one can impose a new parity symmetry which disallows mixing between the ordinary leptons and the new vector lepton fields, under which all the new fields are odd, while the ordinary  leptons are even   \cite{Ishiwata:2013gma}, i.e.,  such that  $  \lambda_E^i= \lambda_L^i =\lambda_{ij}^\prime=\lambda_{ij}^\prime=\lambda_{ij}^{\prime \prime}=0$; alternatively one might choose these couplings to be very small. 
\end{enumerate}

In this analysis the focus will be on Higgs decays.  We investigate the model subjected to  symmetry conditions 1 and/or 2; and we neglect mixing between the ordinary and the new vector leptons. When allowed,  stringent constraints exist on the masses and couplings with ordinary leptons. New vector leptons are ruled out when they mediate flavor-changing neutral currents processes, generate SM neutrino masses or contribute to neutrinoless double beta decay.  Recent studies of models which allow mixing between the ordinary leptons and the new ones exist \cite{Ishiwata:2013gma,Dermisek:2013gta}, but there restrictions from lepton-flavor violating decays force the new leptons to be very heavy $M_L, M_E \sim 10-100$ TeV, or reduces the branching ratio for $h \to \tau^+\tau^-,\, \mu^+ \mu^-$ and of  $h \to \gamma \gamma$ decay to 30-40\% of the SM prediction, neither desirable features for our purpose here. In the Higgs Triplet Model, distinguishing signals would be provided by lighter vector leptons. Since imposing no mixing between ordinary and new leptons allows new lepton masses to be as light as $\sim 100$ GeV, perhaps  without a reduction in the Higgs di-photon branching ratio, we investigate this scenario here. 
 
 In the charged sector, the $2 \times 2$ mass matrix ${\cal M}_E$ is defined as \cite{Ishiwata:2011hr,Joglekar:2012vc}
 \begin{eqnarray}
\left ( E_L^\prime~~e_L^{\prime \prime}\right ) \left ({\cal M}_E \right ) \left( \begin{array}{c} e_R^\prime\\E_R^{\prime \prime} \end{array} \right)\,, \quad {\rm with} \quad
{\cal M}_E=\left ( \begin{array}{cc}
m_E^\prime&M_L\\
M_E&m_E^{\prime\prime}\end{array} \right),  
\end{eqnarray}
with $m_E^\prime=h_E^\prime v_\Phi/\sqrt{2}$ and $m_E^{\prime \prime}=h_E^{\prime \prime} v_\Phi/\sqrt{2}$, with $v_\Phi$ the VEV of the neutral component of the Higgs doublet. The mass matrix can be diagonalized as follows:
\begin{equation}
V_L^\dagger {\cal M}_E V_R=\left ( \begin{array}{cc}
M_{E_1}&0\\
0&M_{E_2} \end{array} \right).
\end{equation}
The mass eigenvalues are
\begin{equation}
M^2_{E_1, E_2}= \frac12 \left [ \left (M_L^2+m_E^{\prime\,2}+M_E^2+m_E^{\prime \prime\,2}\right ) \pm \sqrt{X^2+Y^2} \right ], 
\label{eq:eigenvalues}
\end{equation}
with
\begin{eqnarray}
X&=&\left (M_L^2+m_E^{\prime\,2}-M_E^2-m_E^{\prime \prime\,2}\right ), \nonumber\\
Y&=&2(m_E^{\prime \prime}M_L+m_E^{\prime}M_E).
\end{eqnarray}
By convention, $M_{E_1}>M_{E_2}$. For simplicity we assume lepton Yukawa couplings $h^\prime_E$ and $h^{\prime \prime}_E$ are real so that the transformations that diagonalize the mass matrix are real orthogonal matrices:
\begin{eqnarray}
V_L&=&\left ( \begin{array}{cc}
\cos \theta_L&\sin \theta_L\\
-\sin \theta_L&cos \theta_L \end{array} \right), \\
V_R&=&\left ( \begin{array}{cc}
\cos \theta_R&\sin \theta_R\\
-\sin \theta_R&cos \theta_R \end{array} \right).
\end{eqnarray}
The angles $\theta_{L,R}$ are given by
\begin{eqnarray}
\tan \theta_L=\frac{m_E^{\prime \prime}M_L+m_E^\prime M_E}{M_{E_2}^2-M_L^2-m_E^{\prime\,2}}, \\
\tan \theta_R=\frac{m_E^{\prime }M_L+m_E^{\prime \prime} M_E}{M_{E_2}^2-M_L^2-m_E^{\prime \prime\,2}}.
\end{eqnarray}
The eigenstates of the vector lepton mixing matrix enter in the evaluation of $h \to \gamma \gamma$ and $h \to Z \gamma$ in the next section.
\section{ Production and Decays of the Neutral Higgs Boson}
\label{sec:neutral}

 The presence of the vector leptons affects the loop-dominated decays of the neutral Higgs, $h \to \gamma \gamma$ and $h \to Z \gamma$, and possible  relationships between them. In the Higgs Triplet Model, singly- and doubly-charged bosons also enter in the loops, creating a different dynamic than in the SM. We analyze these decays in turn, and look for possible correlations between them.

\subsection{$h \to \gamma \gamma$}
\label{subsec:gamgam}
Recently, the Triplet Higgs Model has received renewed interest recently because of attempts to reconcile the excess of events in   $h \to \gamma \gamma$ observed at the LHC over those predicted by the SM. Such an enhancement hints at the presence of additional particles, singlets under $SU(3)_c$, but charged under $U(1)_{\rm em}$ which affect only the loop-dominated decay branching ratio, while leaving the production cross section and tree level decays largely unchanged.  Vector leptons are prime candidates for such particles, so we study their contribution to the Higgs decay branching.  The decay width $h \to \gamma \gamma$ is
\begin{eqnarray}
\label{eq:THM-h2gaga}
[\Gamma(h \rightarrow\gamma\gamma)]_{HTM}
& = & \frac{G_F\alpha^2 m_{h}^3}
{128\sqrt{2}\pi^3} \bigg| \sum_{f} N^f_c Q_f^2 g_{h ff} 
A_{1/2}
(\tau^h_f) + g_{h WW} A_1 (\tau^h_W) + \tilde{g}_{h H^\pm\,H^\mp}
A_0(\tau^h_{H^{\pm}}) \nonumber \\
&+& 
 4 \tilde{g}_{h H^{\pm\pm}H^{\mp\mp}}
A_0(\tau^h_{H^{\pm\pm}})   +\frac{\mu_{E_1}g_{hff}}{M_{E_1}} A_{1/2}(\tau^h_{E_1}) +\frac{\mu_{E_2}g_{hff}}{M_{E_2}} A_{1/2}(\tau^h_{E_2})       \bigg|^2 \, ,
\label{partial_width_htm}
\end{eqnarray}
with $M_{E_1}, M_{E_2}$ given in Eq. (\ref{eq:eigenvalues}), and where the loop functions for spin $0$, spin $1/2$ and spin $1$ are given by:
\begin{eqnarray}
A_{0}(\tau) &=& -[\tau -f(\tau)]\, \tau^{-2} \, ,
\label{eq:Ascalar}\\
A_{1/2}(\tau)&=&-\tau^{-1}\left[1+\left(1-\tau^{-1}\right)f\left(\tau^{-1}\right)\right], \label{eq:Afermion}\\
A_1(\tau)&=& 1+\frac32 \tau^{-1}+4 \tau^{-1}\left(1-\frac12 \tau^{-1}\right)f\left(\tau^{-1}\right), \label{eq:Avector} 
\end{eqnarray}
 and the function $f(\tau)$ is given by :
\begin{eqnarray}
f(\tau)=\left\{
\begin{array}{ll}  \displaystyle
\arcsin^2\sqrt{\tau} &0< \tau\leq 1 \\
\displaystyle -\frac{1}{4}\left[ \log\frac{1+\sqrt{1-\tau^{-1}}}
{1-\sqrt{1-\tau^{-1}}}-i\pi \right]^2 \hspace{0.5cm} & \tau>1 \, ,
\end{array} \right. 
\label{eq:ftau} 
\end{eqnarray}
with $\displaystyle \tau^h_i=\frac{m_h^2}{4m_i^2}$, and $m_i$ the mass of the particle running in the loop \cite{Joglekar:2012vc}.  In Eq. (\ref{eq:THM-h2gaga}) the first contribution is from the top quark, the second from the $W$ boson, the third from the singly-charged Higgs boson, the fourth from the doubly-charged Higgs boson, and the last two from the vector leptons. We used the following expressions for the couplings of the Higgs bosons with charged vector leptons:
\begin{eqnarray}
\mu_{E_1}&=&-\cos \theta_L \cos \theta_R \left (m_E^\prime \tan \theta_R+ m_E^{\prime \prime}\tan  \theta_L \right )\nonumber \\
\mu_{E_2}&=&\cos \theta_L \cos \theta_R \left (m_E^\prime \tan \theta_L+ m_E^{\prime \prime}\tan  \theta_R \right )
\end{eqnarray}
The couplings of $h$ to the vector bosons and fermions are as follows:
\begin{eqnarray}
g_{h ff }=\cos\alpha/\cos\beta_{\pm} \, 
\label{littleh1tt};  \qquad 
g_{hWW}= \cos\alpha+2\sin\alpha\, v_\Delta/v_\Phi  \, ,  
\label{littleh1WW} 
\end{eqnarray}
with $ff= t\overline t ,E_1 \overline{E}_1,  E_2 \overline{E}_2$, and the scalar trilinear couplings are parametrized as follows
\begin{eqnarray}
&&\tilde{g}_{h H^{++}H^{--}}  =   \frac{m_W}{ g m_{H^{\pm \pm}}^2} \bigg[ 2\lambda_2v_\Delta \sin\alpha+\lambda_4v_\Phi \cos \alpha \bigg] \, ,\nonumber \\
&&\tilde{g}_{h H^+H^-}= \frac{m_W}{2g m_{H^{\pm}}^2} 
\bigg\{\left[4v_\Delta(\lambda_2 + \lambda_3) \cos^2{\beta_{\pm}}+2v_\Delta\lambda_4\sin^2{\beta_{\pm}}-
\sqrt{2}\lambda_5v_\Phi  \cos{\beta_{\pm}}\sin{\beta_{\pm}}\right]\sin\alpha  \nonumber \\ 
&&+ \left[4\lambda_1\,v_\Phi  \sin{\beta_{\pm}}^2+{(2\lambda_{4}+\lambda_{5}) }v_\Phi  \cos^2{\beta_{\pm}}+
(4\mu-\sqrt{2}\lambda_5v_\Delta)\cos{\beta_{\pm}}\sin{\beta_{\pm}}\right]\cos\alpha\bigg\}  \, . 
\label{eq:redgcallittlehHp}
\end{eqnarray}
Since the new leptons do not affect the Higgs production channels, the effect on the di-photon search channel at the LHC is expressed  by the ratio
\begin{eqnarray}
\label{eq:Rgamgam}
R_{h \to \gamma \gamma}&\equiv& \frac {\sigma_{\rm HTM}(gg \to h \to \gamma \gamma)}{\sigma_{\rm SM}(gg \to  \Phi \to \gamma \gamma)} =\frac{[\sigma( gg \to  h) \times BR(h \to \gamma \gamma)]_{HTM}}{[\sigma( gg \to  \Phi) \times BR(\Phi \to \gamma \gamma )]_{SM}} \nonumber \\
&=&\frac{[\sigma( gg \to  h) \times \Gamma(h \to \gamma \gamma)]_{HTM}}{[\sigma( gg \to\Phi) \times \Gamma(\Phi \to \gamma \gamma)]_{SM}} \times \frac{[\Gamma( \Phi)]_{SM}}{[\Gamma(h)]_{HTM}}, 
\end{eqnarray}
where $\Phi$ is the SM neutral Higgs boson and where the ratio of the cross sections by gluon fusion is
\begin{equation}
\frac {\sigma_{\rm HTM}(gg \to h \to \gamma \gamma)}{\sigma_{\rm SM}(gg \to  \Phi \to \gamma \gamma)}=\cos^2\alpha\,.
\end{equation}
Here $\alpha$ is the mixing angle in the CP-even neutral sector:
\begin{eqnarray}
\left(
\begin{array}{c}
\varphi\\
\delta
\end{array}\right)&=&
\left(
\begin{array}{cc}
\cos \alpha & -\sin\alpha \\
\sin\alpha   & \cos\alpha
\end{array}
\right)
\left(
\begin{array}{c}
h\\
H
\end{array}\right),
\end{eqnarray}
with  
\begin{eqnarray}
\tan2\alpha &=&\frac{v_\Delta}{v_\Phi}\frac{2v_\Phi^2(\lambda_4+\lambda_5)-4{v_\Phi^2\mu}/{\sqrt{2}v_\Delta}}{2v_\Phi^2\lambda_1-{v_\Phi^2\mu}/{\sqrt{2}v_\Delta}-2 v_\Delta^2(\lambda_2+\lambda_3)}.~~~~~~~~ \label{tan2a}
\end{eqnarray}
In \cite{Arbabifar:2012bd} we 
investigated  the Higgs boson decay branching ratio into $\gamma \gamma$ with respect to the SM considering the lightest Higgs boson is the $2.3 \sigma$ signal excess observed at the LEP at 98 GeV, while the heavier Higgs boson is the boson observed at the LHC at 125 GeV, in a Higgs Triplet Model without vector leptons, and found that this is the only scenario which allows for enhancement of the $h \to \gamma \gamma$ branching fraction. We thus set the values 125 GeV and 98 GeV for the $h$ and $H$ masses respectively, and adjust the parameters $\lambda_1-\lambda_5$ accordingly.

\begin{figure}[t]
\center
\vskip -0.3in 
\begin{center}
$
	\begin{array}{ccc}
	\hspace*{-1.2cm}
	\includegraphics[width=3.1in,height=3.0in]{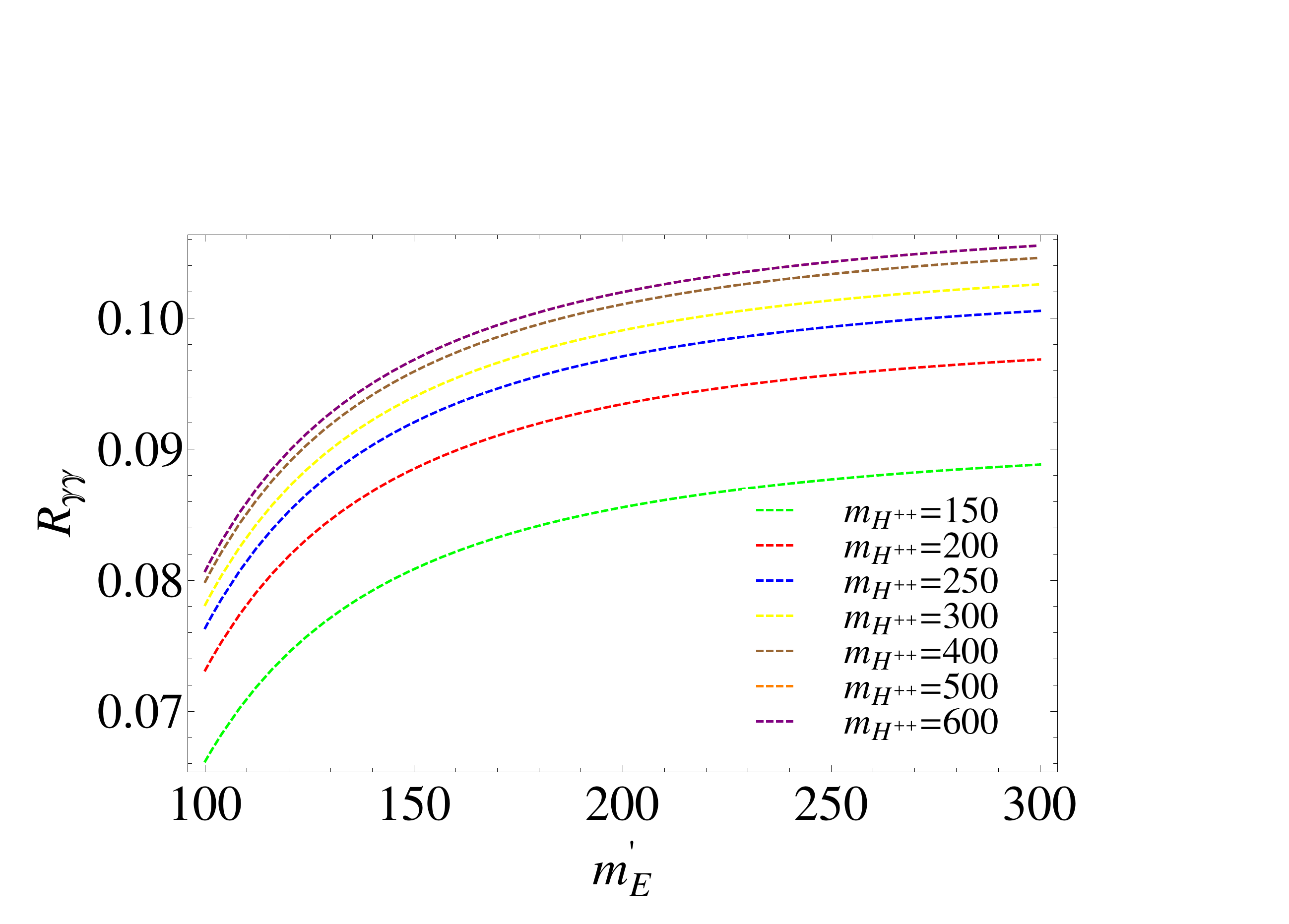}
&\hspace*{-1.5cm}
	\includegraphics[width=3.1in,height=3.1in]{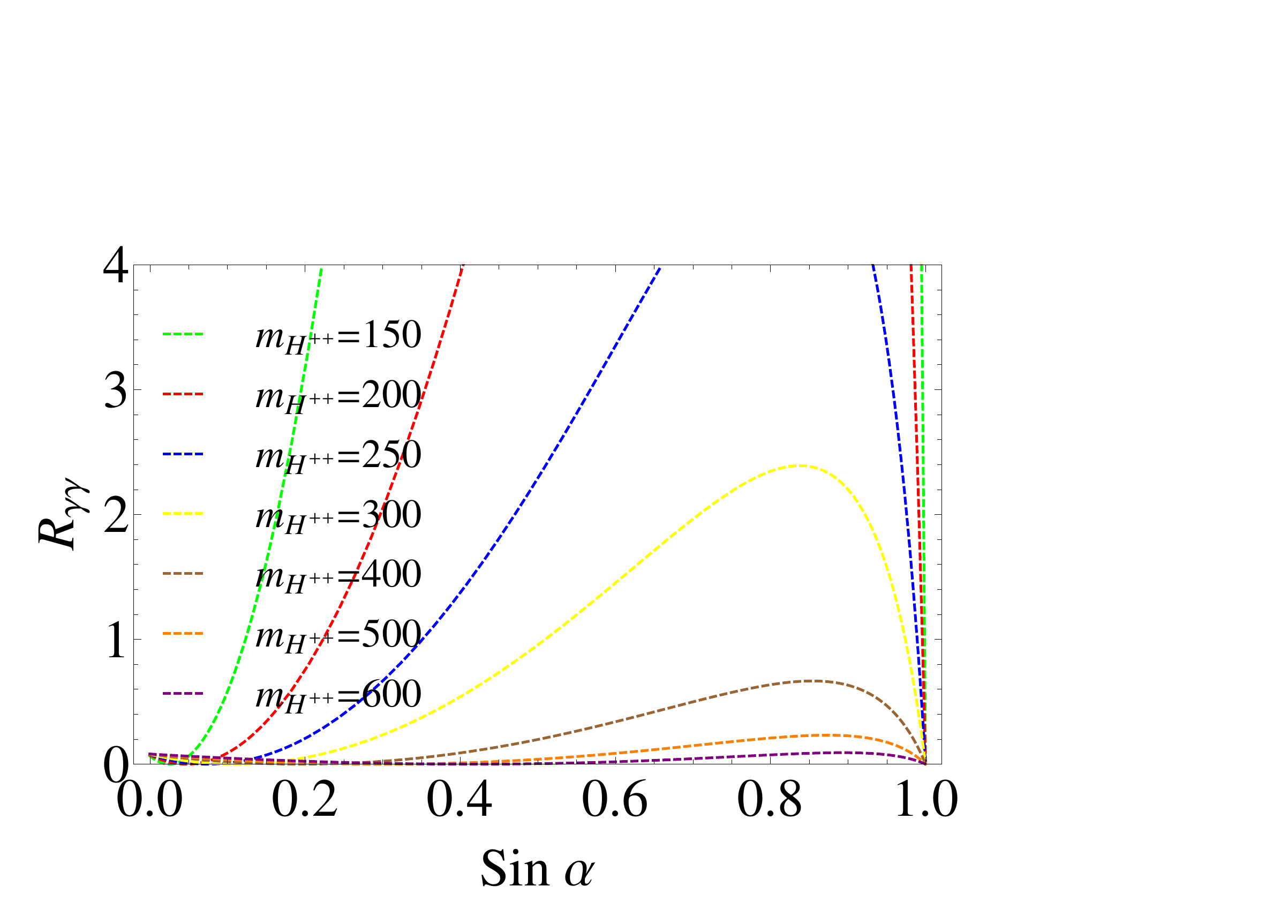}
&\hspace*{-2.1cm}
	\includegraphics[width=3.1in,height=3.1in]{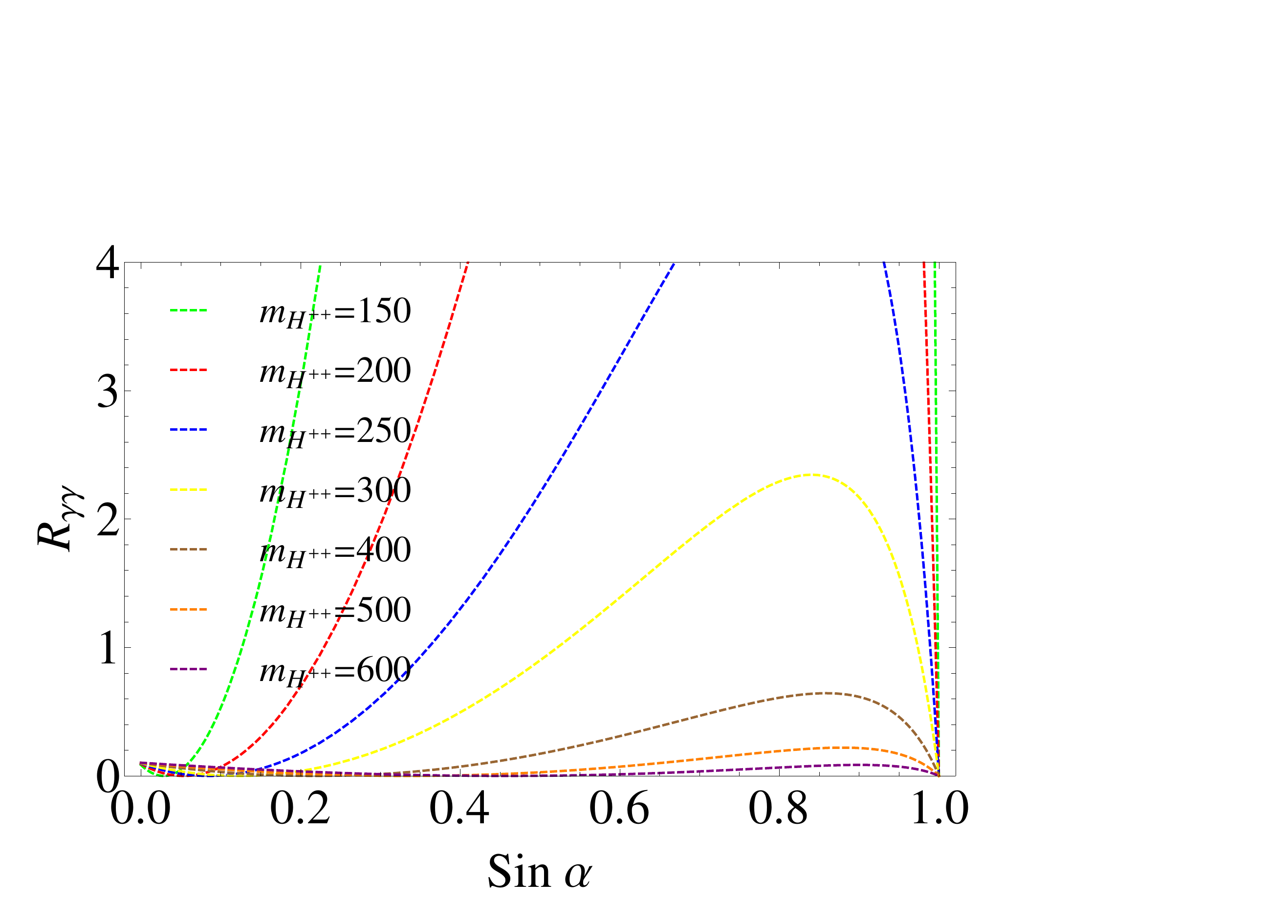}
        \end{array}$
        \end{center}
\caption{Relative decay rate $R_{h \to \gamma \gamma}$  in the limit $M_L=M_E=0$ (left panel) as a function of $m_E^{\prime} =m_E^{\prime \prime}$, for $\sin \alpha=0$; (middle panel) as a function of $\sin \alpha$  for $m_E^{\prime} =m_E^{\prime \prime}=100$ GeV, and (right panel) as a function of $\sin \alpha$ for $m_E^{\prime} =m_E^{\prime \prime}=200$ GeV. The colored-coded  curves correspond to  different values of doubly-charged Higgs masses, given in the attached panels in GeV. } 
\label{fig:hgg}
\end{figure}

Vector lepton masses and mixing parameters depend on $M_L$ and $M_E$, the explicit mass parameters in the Lagrangian; and $h_E^\prime, h_E^{\prime \prime}$, the vector leptons Yukawa parameters. 
 In the limit of vanishing Dirac mass terms $M_L$ and $M_E$, the pre-factors $\displaystyle \frac{\mu_{E_i}}{M_{E_i}}$  in (11) go to one. It then follows that there is destructive interference between the dominant $W$- boson contribution and the charged leptons loops \cite{Joglekar:2012vc}.  In this limit, despite possible enhancement from singly- and doubly-charged Higgs bosons in the loop, we find a large suppression of the di-photon rate. We present the plots for the relative signal strength  of $R_{h \to \gamma \gamma}$, defined in Eq. (\ref{eq:Rgamgam}) as a function of $m_E^\prime =m_E^{\prime \prime}$ (or equivalently $h_E^\prime=h_E^{\prime \prime}$), for various values of doubly charged Higgs bosons mass, in the left-side panel of Fig. \ref{fig:hgg}, for $\sin \alpha = 0$.  Clearly,  for this case (no mixing) the decay of the $h$ is suppressed significantly with respect to the value in SM over the whole region of the parameter space in $m_E^\prime$. 

Allowing mixing in the neutral Higgs sector changes the relative contributions of the charged Higgs to the di-photon decay. We show decay rates for $h \to \gamma \gamma$ as a function of $\sin \alpha$ for different values of doubly-charged Higgs boson mass considering $m_E^\prime=m_E^{\prime \prime}=100$ GeV (and 200 GeV)  in Fig.  \ref{fig:hgg} middle (and right) panels respectively. Considerations for relative branching ratios are affected by the fact that the total width of Higgs boson in the HTM for $\sin \alpha \ne 0$ is not the same as in the SM. The relative widths factor is
\begin{eqnarray}
\label{eq:width}
\!\!\!\!\!\!\frac{[\Gamma(h)]_{HTM}}{[\Gamma(\Phi)]_{SM}}=\frac{[\Gamma (h \to \sum\limits_f f {\bar f})+ \Gamma (h \to WW^*) +\Gamma (h \to ZZ^*)+ \Gamma (h \to \nu \nu)]_{HTM}}{[\Gamma (\Phi \to \sum\limits_f f {\bar f})+ \Gamma ( \Phi \to WW^*) +\Gamma ( \Phi \to ZZ^*)]_{SM}} . 
\end{eqnarray}
The plots in Fig. \ref{fig:hgg} correspond to symmetry condition 1 in Section \ref{sec:model}, that is, $M_L=M_E=0$.

\begin{figure}[t]
\center
\begin{center}
$
	\begin{array}{ccc}
\hspace*{-0.9cm}
\includegraphics[width=2.3in,height=2.2in]{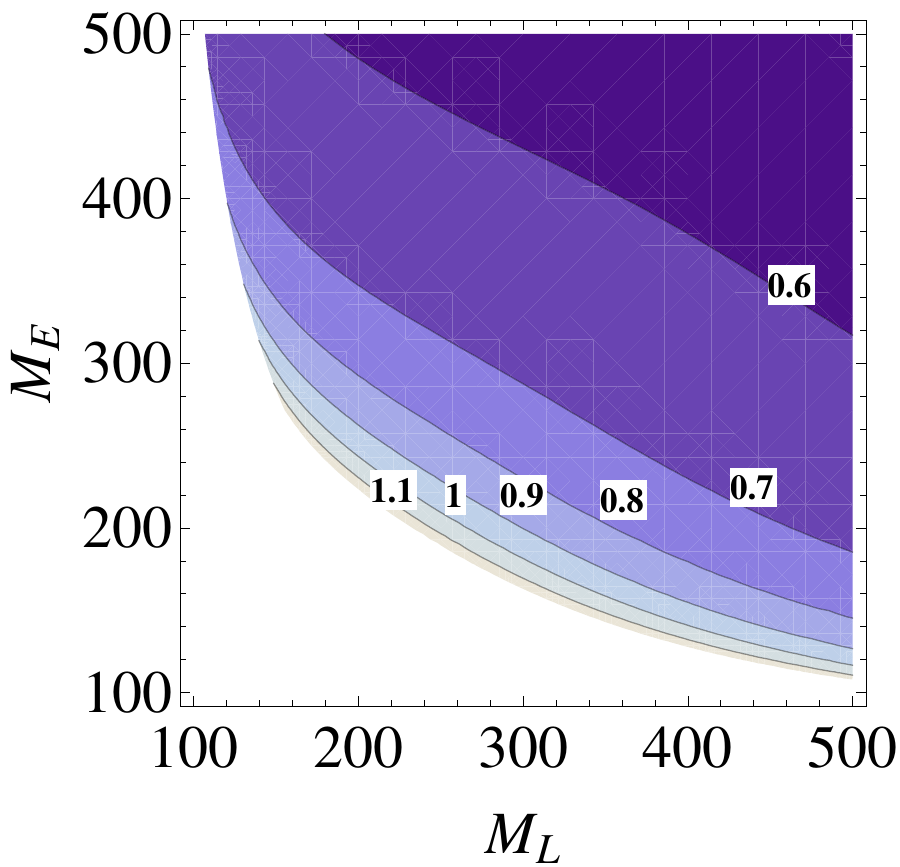}
&\hspace*{-0.4cm}
	\includegraphics[width=2.3in,height=2.2in]{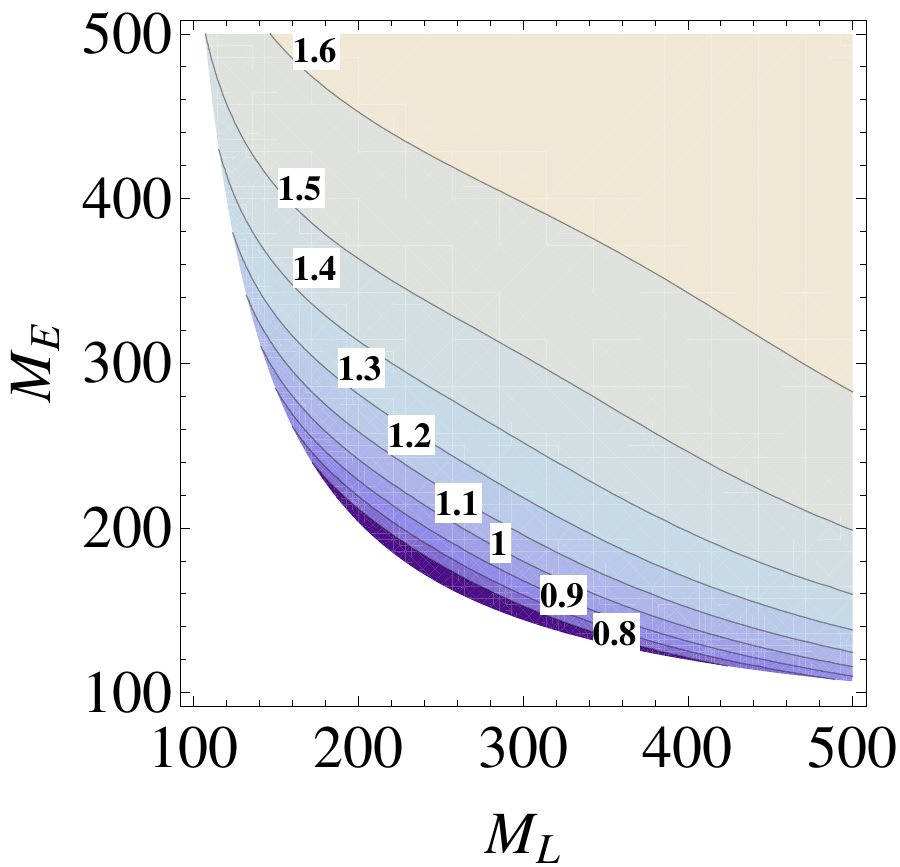}
&	\hspace*{-0.4cm}
	\includegraphics[width=2.3in, height=2.2in]{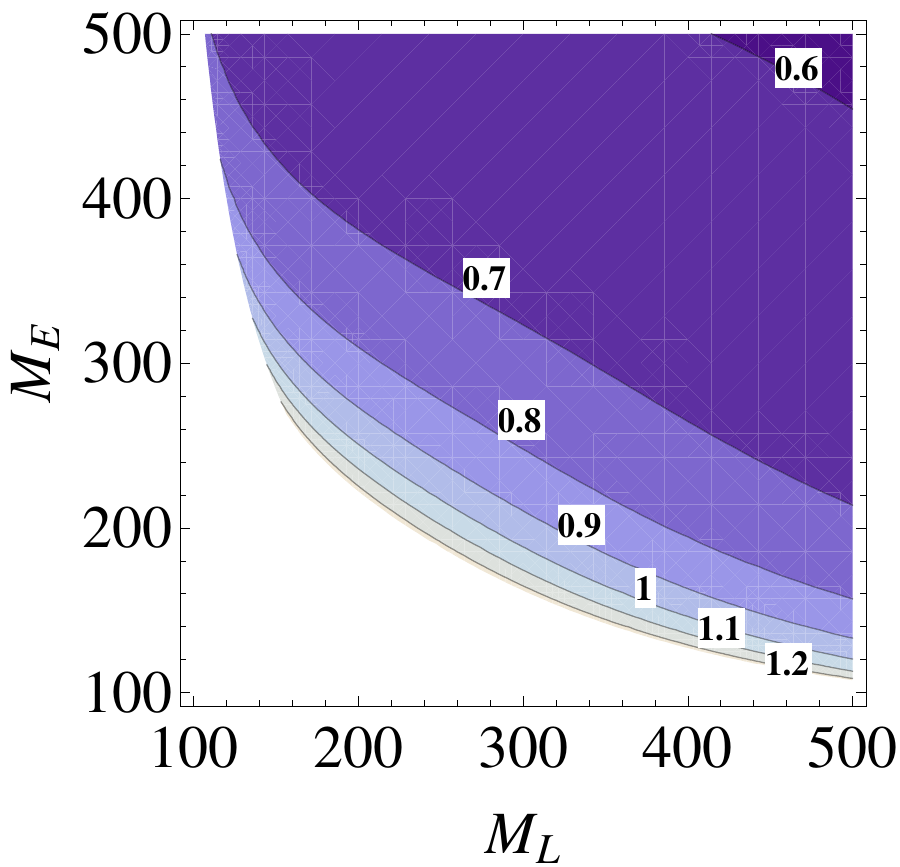}\\
\hspace*{-0.9cm}
	\includegraphics[width=2.3in,height=2.2in]{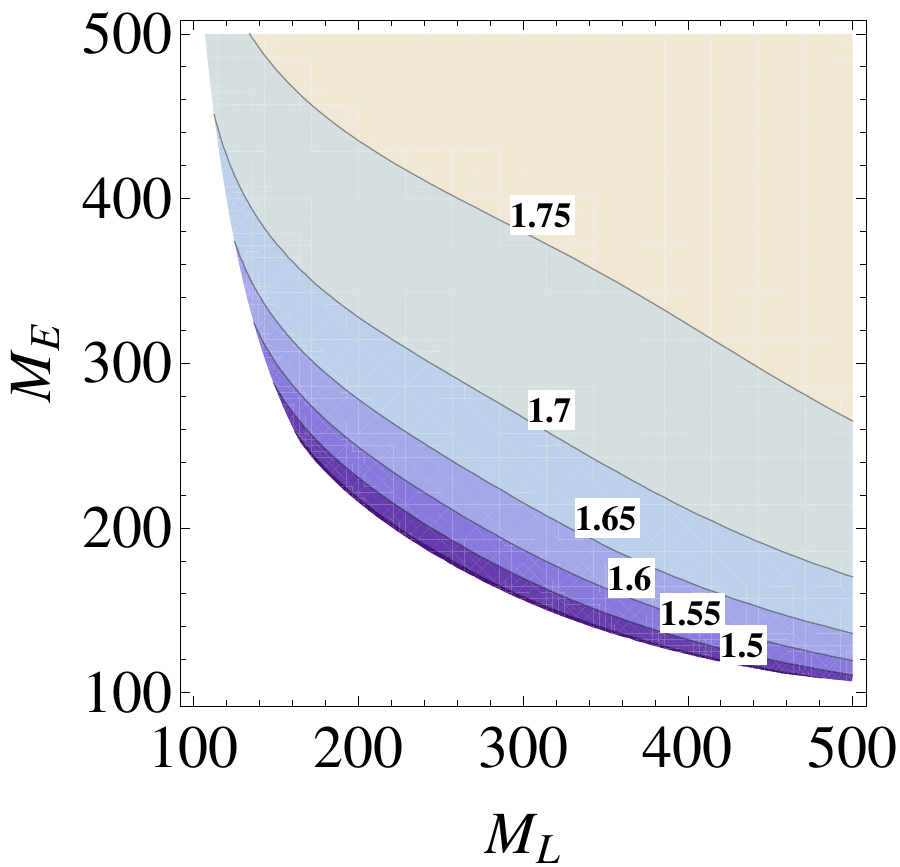}
	&\hspace*{-0.4cm}
\includegraphics[width=2.3in,height=2.2in]{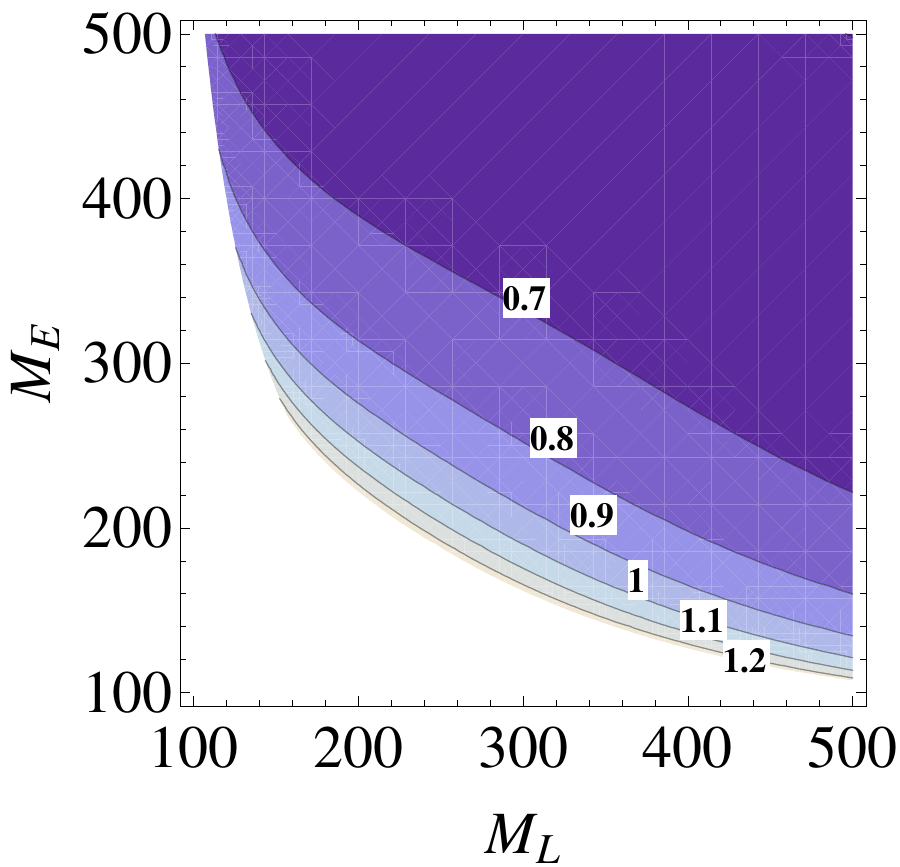}
&\hspace*{-0.4cm}
	\includegraphics[width=2.3in,height=2.2in]{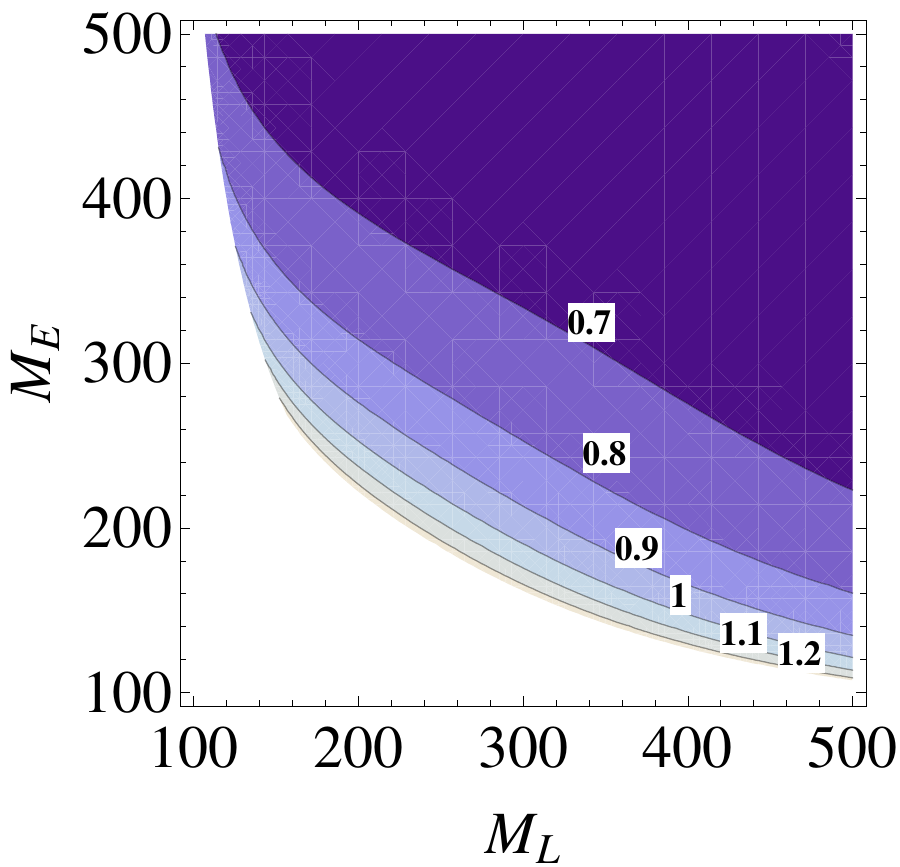}
        \end{array}$
        \end{center}
\caption{Contour plots  of constant $R_{h\to \gamma \gamma}$ for mass terms $M_E$ and $M_L$, for $h_E^{\prime}= h_E^{\prime \prime}=0.8$ and  combinations of doubly-charged Higgs boson masses and $\sin \alpha$: (upper left panel) $m_{H^{\pm \pm}} =150$ GeV, $\sin \alpha=0$; (upper middle panel)  $m_{H^{\pm \pm}} =150$ GeV, $\sin \alpha=0.2$;  (upper right panel) $m_{H^{\pm \pm}} =300$ GeV, $\sin \alpha=0$ ; and (lower left panel) $m_{H^{\pm \pm}} =300$ GeV, $\sin \alpha=0.9$ ; (lower middle panel) $m_{H^{\pm \pm}} =500$ GeV, $\sin \alpha=0$;   (lower right  panel) $m_{H^{\pm \pm}} =600$ GeV, $\sin \alpha=0$. }  
\label{fig:hEmLcont1}
\end{figure}

\begin{figure}[t]
\center
\begin{center}
$
	\begin{array}{ccc}
\hspace*{-0.9cm}
\includegraphics[width=2.3in,height=2.2in]{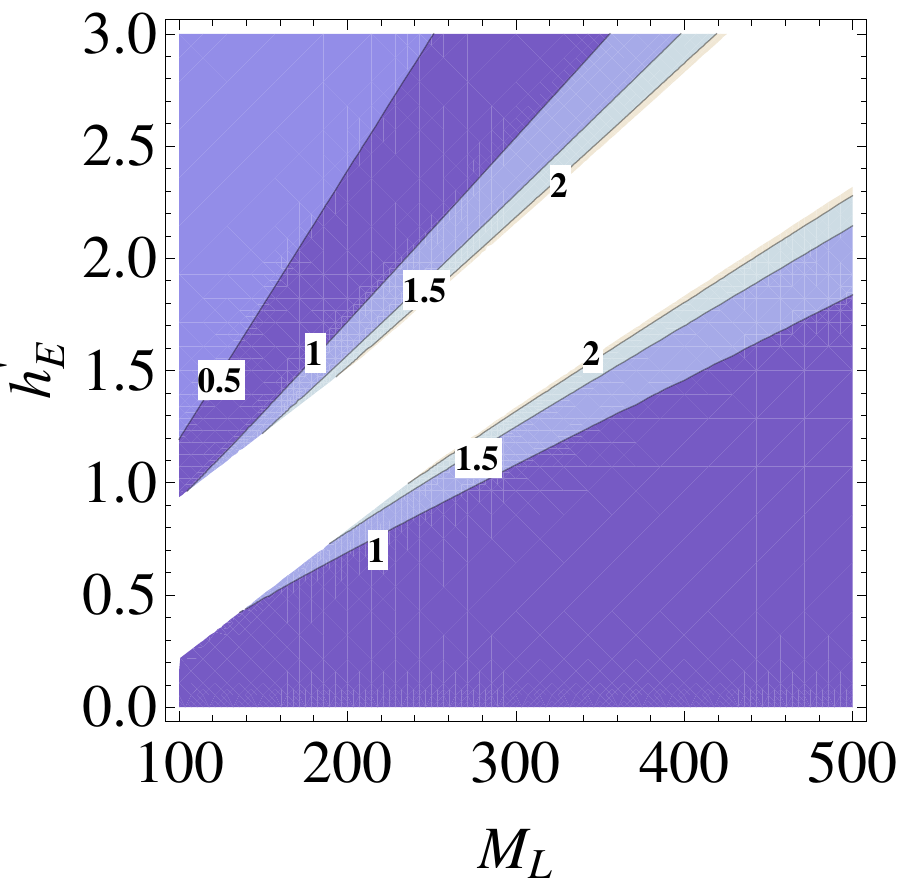}
&\hspace*{-0.4cm}
	\includegraphics[width=2.3in,height=2.2in]{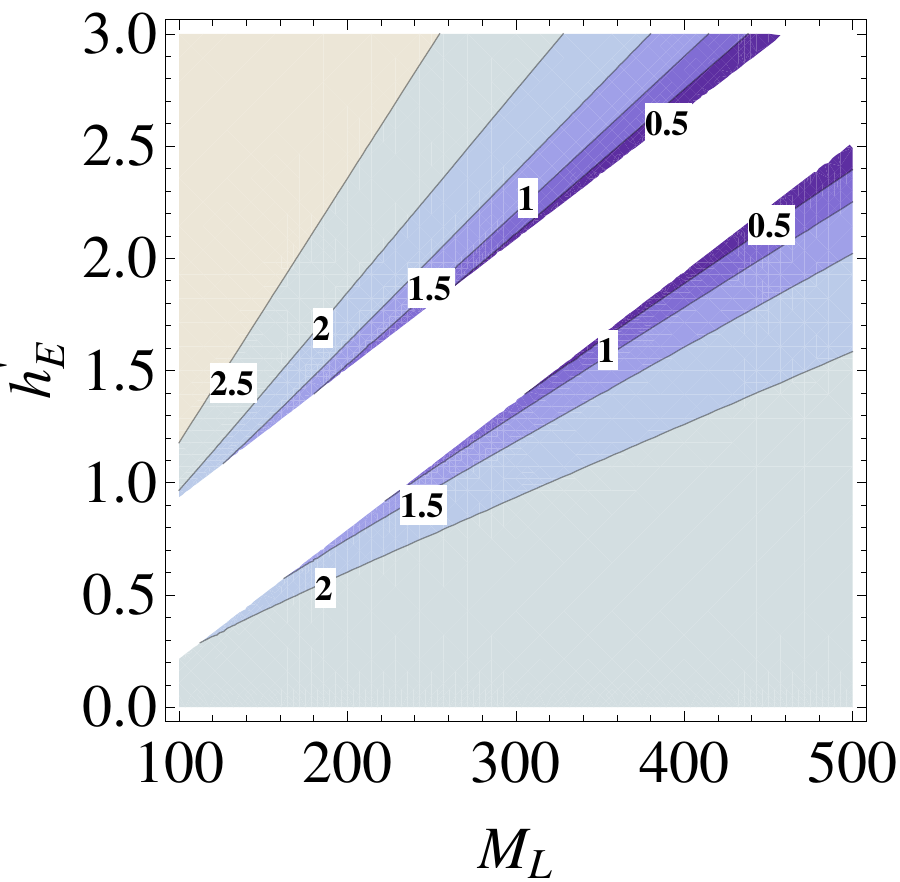}
&	\hspace*{-0.4cm}
	\includegraphics[width=2.3in, height=2.2in]{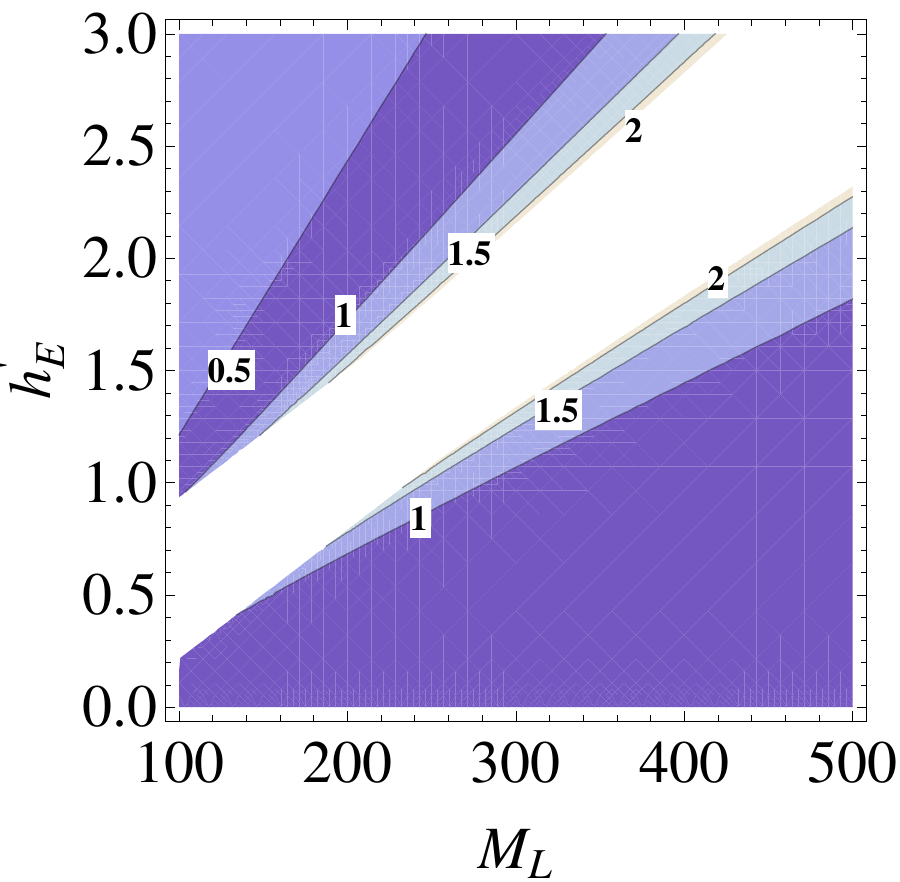}\\
\hspace*{-0.9cm}
	\includegraphics[width=2.3in,height=2.2in]{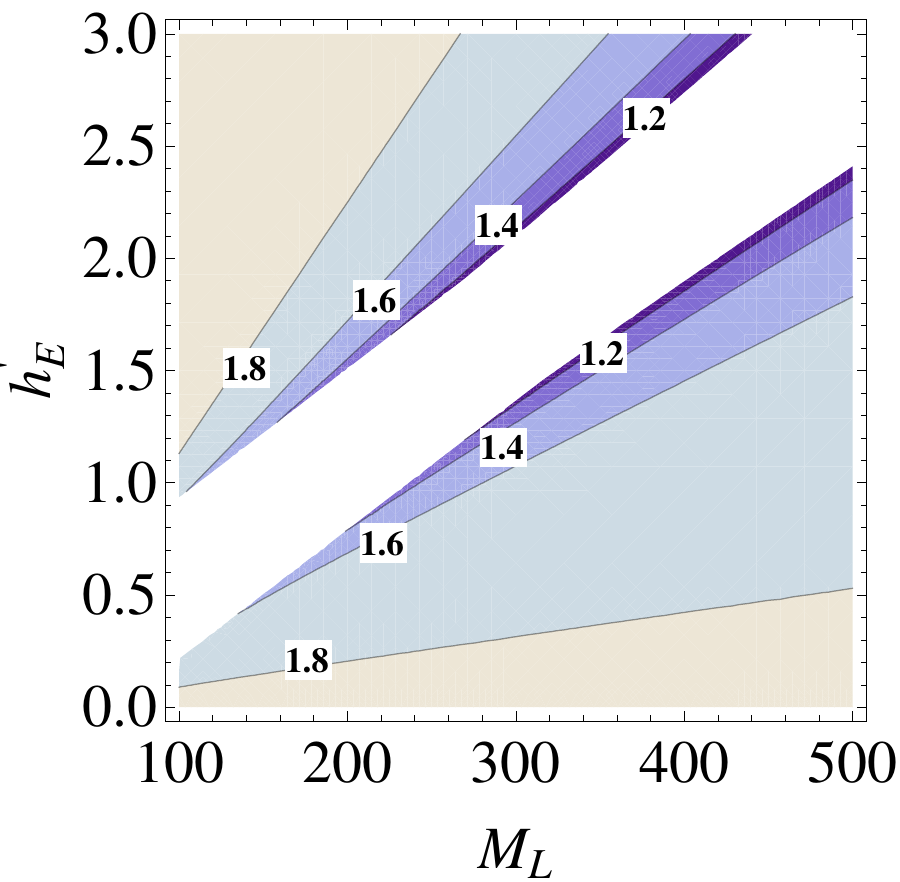}
	&\hspace*{-0.4cm}
\includegraphics[width=2.3in,height=2.2in]{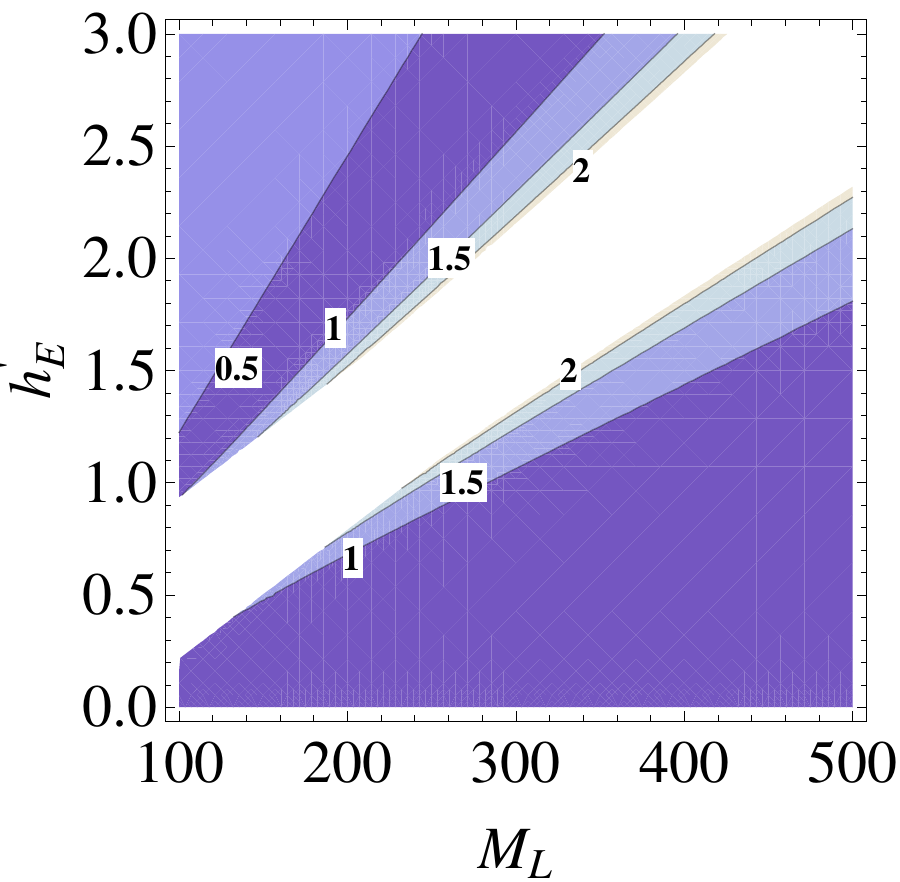}
&\hspace*{-0.4cm}
	\includegraphics[width=2.3in,height=2.2in]{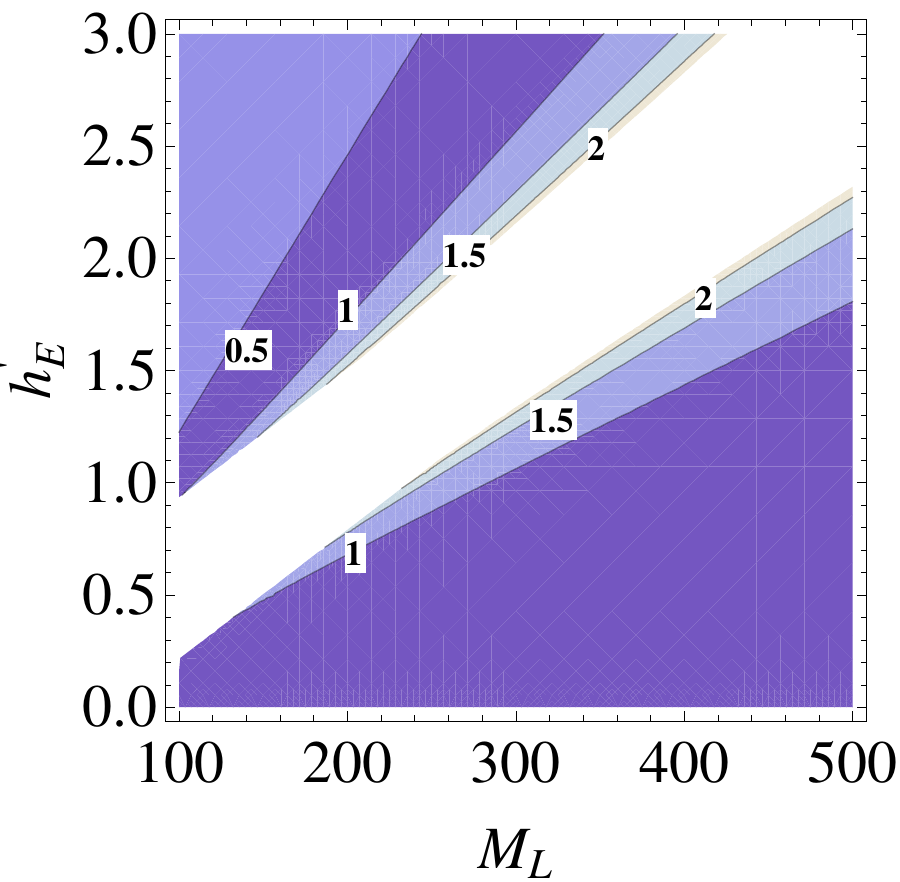}
        \end{array}$
        \end{center}
\caption{ Contour plots of constant $R_{h\to \gamma \gamma}$ for mass terms  $M_L$ and  $h_E^\prime=h_E^{\prime \prime}$,  for various doubly-charged Higgs boson masses and $\sin \alpha$:  (upper left panel) $m_{H^{\pm \pm}} =200$ GeV, $\sin \alpha=0$;   (upper middle panel) $m_{H^{\pm \pm}} =200$ GeV, $\sin \alpha=0.4$;   (upper right panel) $m_{H^{\pm \pm}} =300$ GeV, $\sin \alpha=0$; and  (lower left panel) $m_{H^{\pm \pm}} =300$ GeV, $\sin \alpha=0.9$;  (lower middle panel) $m_{H^{\pm \pm}} =500$ GeV, $\sin \alpha=0$;  (lower right panel) $m_{H^{\pm \pm}} =600$ GeV, $\sin \alpha=0$.} 
\label{fig:hEmLcont2}
\end{figure}

However, if mixing with SM leptons is forbidden, but the vector leptons are still allowed to mix with each other,  the pre-factors $\mu_{E_i }/M_{E_i}$  in Eq. (\ref{eq:THM-h2gaga}) are not one, and can modify the Higgs di-photon decay. In the next plots we investigate the effect of non-zero mass parameters $M_L$ and $M_E$, for fixed values of the Yukawa couplings. 
In Fig.  \ref{fig:hEmLcont1} we present the contour plots of constant $R_{h\to \gamma \gamma}$ for $h_E^{\prime}= h_E^{\prime \prime}=0.8$  in the plane of the explicit mass terms $M_L $ and $M_E$, for various values of doubly-charged Higgs bosons mass and $\sin \alpha $. The contours are labeled by the value of $R_{h\to \gamma \gamma}$. The vector lepton masses are restricted to values for which \cite{Ishiwata:2011hr}
$
M_{E_2} \ge 62.5~{\rm GeV} 
$, 
where $M_{E_{1,2}}$ are given in Eq. (\ref{eq:eigenvalues}). The plots indicate that it is difficult to obtain any significant enhancement of the ratio  $R_{h\to \gamma \gamma}$ for $\sin \alpha=0$, and this does not depend on the chosen values for the doubly-charged Higgs mass; while for $\sin \alpha \ne 0$, enhancements are possible for various values of $m_{H^{\pm \pm}}$. 
In Fig. \ref{fig:hEmLcont2} we investigate the dependence of $R_{h\to \gamma \gamma}$ on the Yukawa couplings and vector lepton masses. We show contour plots for fixed $R_{h \to \gamma \gamma}$ in a $h_E^\prime-M_L$ plane, with $h_E^\prime=h_E^{\prime \prime}$ and $M_L=M_E$, for various values of $\sin \alpha$ and doubly-charged Higgs boson masses. Enhancements are possible here  for all values of $\sin \alpha$, but while for $\sin \alpha=0$ the decay $h \to \gamma \gamma$ is enhanced for large vector lepton masses and Yukawa couplings, for  $\sin \alpha \ne 0$ we observe enhancements for light vector lepton masses and small Yukawa couplings.

If  we wish to study the light vector leptons parameter space where  $h \to \gamma \gamma$ is enhanced,    $\sin \alpha \ne 0$ is preferred.  The enhancement is affected by mixing in the vector lepton sector, the various values for doubly-charged Higgs Bosons mass and values of $\sin \alpha$. 

As the plots cover only  a limited range of the parameter space, in the Tables below we give the ranges for the values of the ratio $R_{h \to \gamma \gamma} $ for the various scenarios. In Table \ref{tab:Y1/6_ME_ML}, we fix the value of the Yukawa coupling to $h^\prime_E=0.8$,  allow the vector lepton masses to vary in the (100-500) GeV range, and show the values for $R_{h \to \gamma \gamma}$  for different  $\sin \alpha$ and doubly-charged Higgs masses.  We note that the relative branching ratios are very sensitive to both doubly charged Higgs mass values and values of $\sin \alpha$. Enhancements in the branching ratio of $h \to \gamma \gamma$ are possible for light $m_{H^{\pm \pm}} \le 300$ GeV, and  are much more pronounced at large $\sin \alpha$. Note that for $\sin \alpha=0$, the result is independent of $m_{H^{\pm \pm}}$, in agreement with the results obtained in \cite{Arbabifar:2012bd}. The reason is the following. In Eq. (\ref{eq:redgcallittlehHp}), for $\sin \alpha=0$,  the coupling between neutral and doubly-charged Higgs 
\begin{eqnarray}
\tilde{g}_{h H^{++}H^{--}}  =   \frac{m_W}{ g m_{H^{\pm \pm}}^2} \bigg[ \lambda_4v_\Phi \bigg] = \frac{m_W}{ g m_{H^{\pm \pm}}^2}\bigg[   2\frac {m^2_{H^{\pm \pm}}}{v_\Phi^2} v_\Phi \bigg]=2 \frac{m_W }{g v_\Phi}, 
\end{eqnarray}
where we used the expression for $\lambda_4$  from \cite{Arbabifar:2012bd}, is independent of $m_{H^{\pm \pm}}$. 
%
\begin{table}[htbp]
\caption{\label{tab:Y1/6_ME_ML}\sl\small Range of the ratio $R_{h \to \gamma \gamma}$, as defined in the text, for  doubly-charged Higgs mass (in columns) and neutral Higgs mixing angle $\sin \alpha$ (in rows),  for Dirac vector lepton masses in the range $M_E, M_L \in (100-500)$ {\rm GeV}, with  $h^\prime_E=0.8$
.}
  \begin{center}
\footnotesize{
 \begin{tabular*}{0.99\textwidth}{@{\extracolsep{\fill}} c|cccccc}
 \hline\hline
  $R_{\gamma \gamma}$ &$m_{H^{\pm \pm}}=150$ GeV &$200$ GeV
 &$250$ GeV &$300$ GeV &$500$ GeV &$600$ GeV\\
  \hline\hline
  $\sin \alpha=0$ &$0.6-1.2$ &$0.6-1.2$ &$0.6-1.2$ &$0.6-1.2$ &$0.6-1.2$ &$0.6-1.2$  \\
  $\sin \alpha=0.1$ &$0.02-0.08$ &$0.05-0.23$ &$0.2-0.5$ &$0.3-0.7$  &$0.5-1$ &$0.6-1$ \\
  $\sin \alpha=0.2$ &$0.8-1.6$ &$0.02-0.14$  &$0.01-0.08$ & $0.1-0.3$  &$0.4-0.8$ &$ 0.5-0.9$  \\
  $\sin \alpha=0.3$ &$4-5.2$ &$0.2-0.9$  &$0.02-0.12$ &$0.01-0.04$  &$0.25-0.55$ &$0.3-0.7$   \\
  $\sin \alpha=0.4$ &$9-10.75$ &$1.4-2.2$  &$0.1-0.5$ &$0.02-0.08$  &$0.15-0.35$ &$0.25-0.55$ \\
  $\sin \alpha=0.5$ &$16-18$ &$3.2-4.2$  &$0.6-1.2$ &$0.05-0.3$  &$0.06-0.22$ &$0.15-0.35$   \\
  $\sin \alpha=0.6$ &$24-26.5$  &$5.6-6.8$  &$1.4-2.1$ &$0.25-0.7$ &$0.01-0.07$ & $0.06-0.2$ \\
  $\sin \alpha=0.7$ &$32.5-34.5$ &$8.2-9.4$  &$2.4-3.1$ &$0.7-1.1$  &$0.002-0.008$ &$0.02-0.08$    \\
  $\sin \alpha=0.8$ &$38.5-40.75$  &$10.4-11.4$  &$3.4-4.1$ &$1.2-1.65$ &$0.005-0.045$ &$0.001-0.006$  \\
  $\sin \alpha=0.9$  &$36.2-37.4$  &$10.2-10.9$ &$3.7-4.1$ &$1.5-1.75$ &$0.04-0.01$ &$0.005-0.002$ \\
    \hline
    \hline
   \end{tabular*}
   }
\end{center}
 \end{table}
In Table \ref{tab:Y1/6_h_E} we  allow, in addition to mass variations,  variations in the Yukawa coupling $h^\prime_E\in (0-3)$. This means allowing both explicit (Dirac) masses and additional contributions by electroweak symmetry breaking, $m_E^\prime,\, m_E^{\prime \prime}$. The dependence on  the Yukawa coupling $h^\prime_E$ is much weaker than on $\sin \alpha$ or on $m_{H^{\pm \pm}}$.


%
\begin{table}[htbp]
\caption{\label{tab:Y1/6_h_E}\sl\small Same as in Table \ref{tab:Y1/6_ME_ML}, but also allowing  $ h^\prime_E \in (0-3)$.}
  \begin{center}
 \footnotesize{
 \begin{tabular*}{0.99\textwidth}{@{\extracolsep{\fill}} c|cccccc}
 \hline\hline
  $R_{\gamma \gamma}$ &$m_{H^{\pm \pm}}=150$ GeV &$200$ GeV
 &$250$ GeV &$300$ GeV &$500$ GeV &$600$ GeV\\
  \hline\hline
  $\sin \alpha=0$ &$0.5-2$ &$0.5-2$ &$0.5-2$ &$0.5-2$ &$0.5-2$ &$0.5-2$  \\
  $\sin \alpha=0.1$ &$0.05-0.15$ &$0.1-0.6$ &$0.2-1$ &$0.2-1.4$  &$0.5-1.75$ &$0.5-2$ \\
  $\sin \alpha=0.2$ &$0.5-2$ &$0.1-0.4$  &$0.05-0.3$ & $0.1-0.7$  &$0.25-1.5$ &$ 0.25-1.75$  \\
  $\sin \alpha=0.3$ &$2-6$ &$0.5-1$  &$0.1-0.4$ &$0.05-0.2$  &$0.2-1.2$ &$0.2-1.4$   \\
  $\sin \alpha=0.4$ &$6-11$ &$1-2.5$  &$0.2-0.6$ &$0.05-0.3$  &$0.2-0.8$ &$0.2-1$ \\
  $\sin \alpha=0.5$ &$14-18$ &$2-4$  &$0.5-1.5$ &$0.2-0.4$  &$0.1-0.5$ &$0.2-0.8$   \\
  $\sin \alpha=0.6$ &$20-26$  &$4-7$  &$0.5-2.5$ &$0.25-0.75$ &$0.05-0.25$ & $0.1-0.5$ \\
  $\sin \alpha=0.7$ &$30-36$ &$7-10$  &$1.5-3.5$ &$0.25-1.25$  &$0.01-0.07$ &$0.05-0.2$    \\
  $\sin \alpha=0.8$ &$36-40$  &$9-12$  &$2.5-4$ &$0.5-1.75$ &$0.02-0.08$ &$0.01-0.04$  \\
  $\sin \alpha=0.9$  &$35-38$  &$9.5-11$ &$3.2-4$ &$1.2-1.8$ &$0.05-0.1$ &$0.04-0.01$ \\
    \hline
    \hline
   \end{tabular*}
   }
\end{center}
 \end{table}
However, one can see from the Tables that modest enhancements of the ratio $R_{h \to \gamma \gamma}$ are possible for $\sin \alpha=0$ for large vector leptons Yukawa couplings, unlike in the case of the Triplet Model without vector leptons. This would then be a clear distinguishing feature: enhancements of the decay $h \to \gamma \gamma$ in the absence of mixing in the neutral sector. The absence of mixing would manifest itself in observing tree-level decays ($ h \to b\bar b, \tau^+\tau^-, ZZ^*$ and $WW^*$) identical to those in the SM. There seems to be a minimum value of $R_{h \to \gamma \gamma}$ for $\sin \alpha=0.1$, where the contribution from the doubly-charged Higgs bosons is important for small doubly-charged masses and counters the contribution from the vector leptons. This is  a suppression of the branching ratio for $h \to \gamma \gamma$  due to the fact that the vector lepton contribution interacts destructively with the dominant $W^\pm$ contribution.  As a general feature,  $R_{h \to \gamma \gamma}$ increases when we lower the doubly charged Higgs mass  and increase $\sin \alpha$. This rules out part of the parameter space. For instance, for $m_{H^{\pm\pm}}=150$ GeV, the mixing cannot be larger than $\sin \alpha=0.2$, and for $m_{H^{\pm \pm}} =200$  GeV, mixings  larger than $\sin \alpha \ge 0.5$ are ruled out.  If the value of $m_{H^{\pm\pm}}$ is increased to $500-600$ GeV, only modest enhancements are possible, and only for $\sin \alpha=0$, for vector lepton explicit masses in the $100-500$ GeV range and $h^\prime_E=0.8$. Increasing the vector leptons Yukawa coupling increases the overall ratio $R_{h \to \gamma \gamma}$.
\subsection{$h \to Z \gamma$}
\label{subsec:Zgam}
\begin{figure}[t]
\center
\vskip -0.3in 
\begin{center}$
	\begin{array}{ccc}
\hspace*{-1.2cm}
\includegraphics[width=3.1in,height=3.0in]{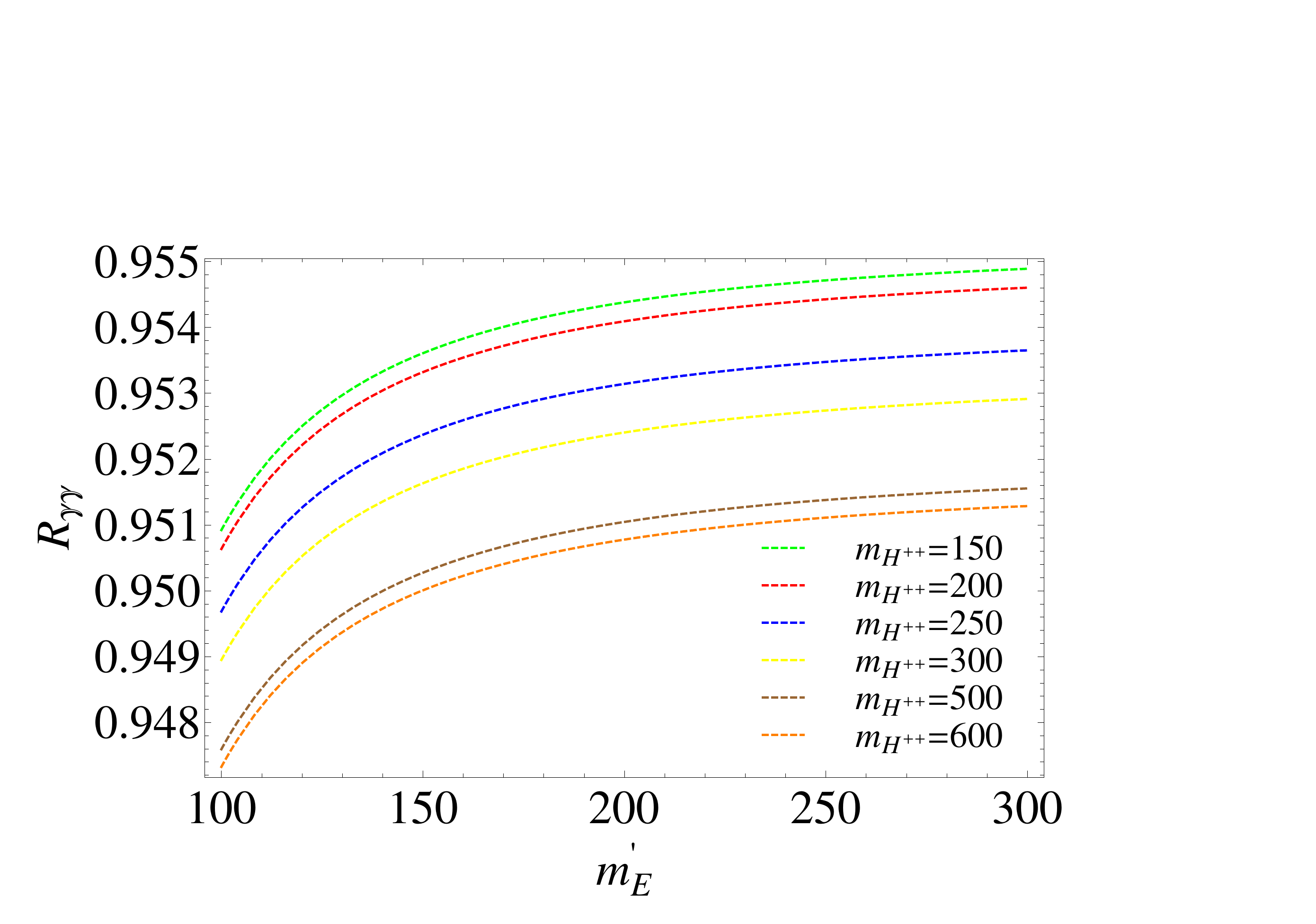}
&\hspace*{-1.7cm}
	\includegraphics[width=3.1in,height=3.0in]{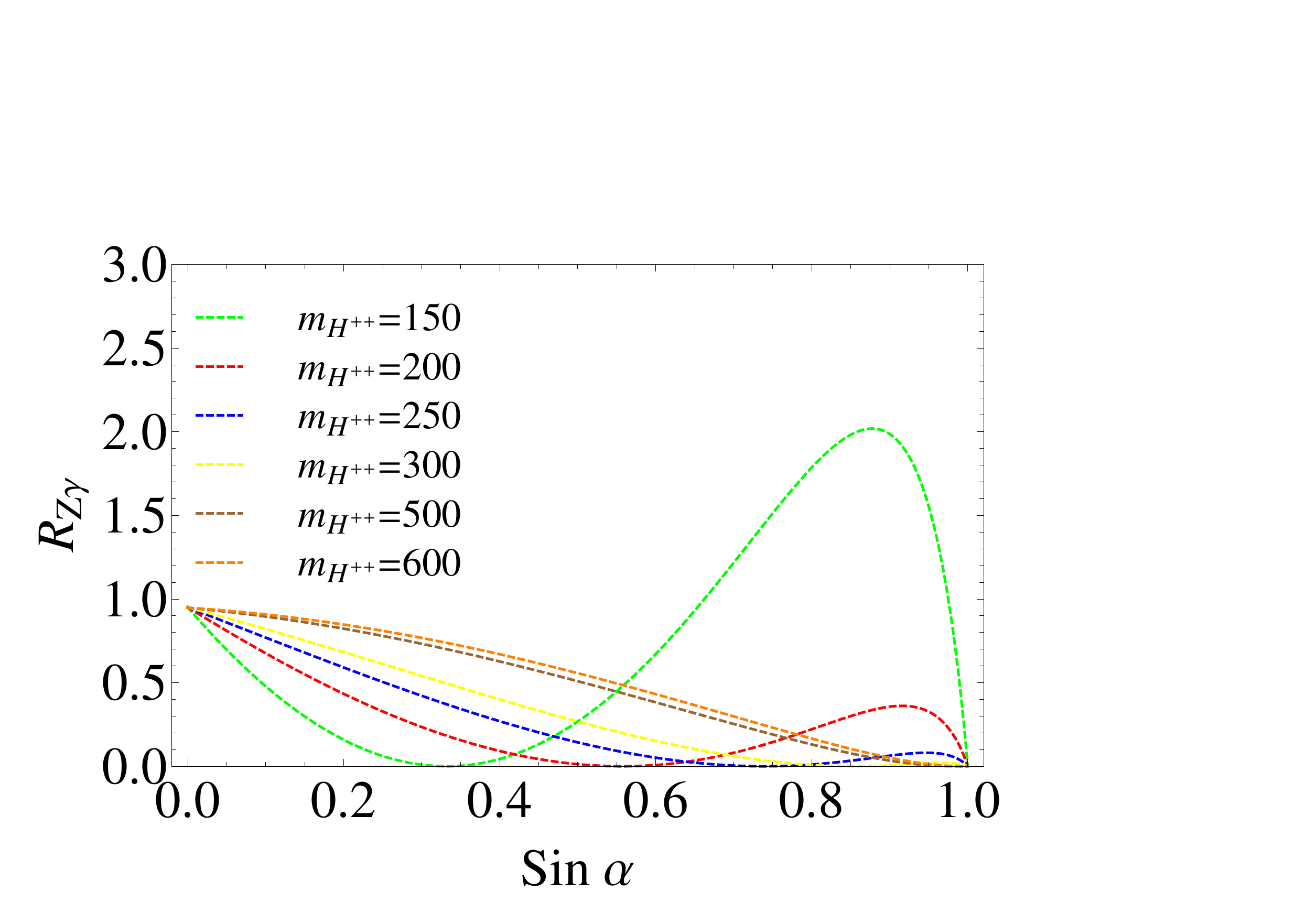}
&	\hspace*{-1.9cm}
	\includegraphics[width=3.1in, height=3.0in]{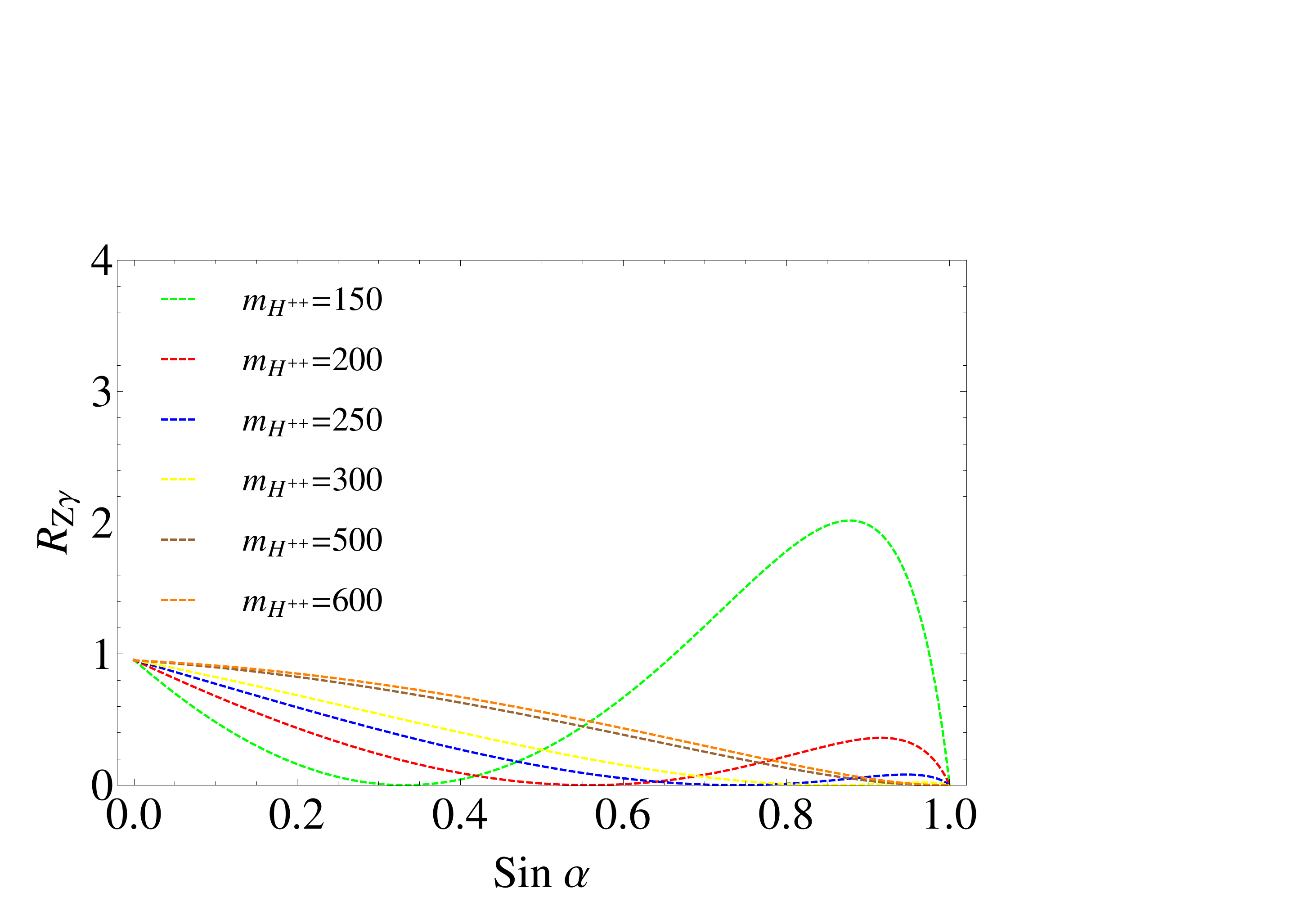}
        \end{array}$
        \end{center}
\caption{Relative decay rate for $R_{h \to \gamma Z}$ as a function of $m_E^{\prime} =m_E^{\prime \prime}$ for different values of doubly-charged Higgs masses, in the case of no mixing, i.e., $\sin \alpha=0$ (left panel); and as a function of $\sin \alpha$  for $m_E^{\prime} =m_E^{\prime \prime}=100$ GeV (middle panel), and  $m_E^{\prime} =m_E^{\prime \prime}=200$ GeV (right panel). The colored-coded  curves correspond to  different values of doubly-charged Higgs masses, given in the attached panels in GeV.} 
\label{fig:hgZmE}
\end{figure}

In most models, the $h\to \gamma\gamma$ and $h\to Z\gamma$ partial decay widths are correlated or anti-correlated, though usually the enhancement/suppression in the $Z\gamma$ channel is much smaller compared to that in the $\gamma\gamma$ channel. However, as in models with new loop-contributions to $ h \to \gamma \gamma, Z \gamma$, sensitivity to both is expected, we study the correlation between the two here, in the presence of vector leptons. Investigation of the branching ratio of $h \to Z \gamma$ is also further justified by the recent results from CMS and ATLAS \cite{Chatrchyan:2013vaa}, which indicate branching fractions consistent with the SM expectation at 1$\sigma$ in the Higgs boson $h$ mass region at 95\% C.L. .
 The decay width for $h \to Z \gamma$ is given by \cite{Chen:2013vi}:
\begin{eqnarray}
\label{eq:hzg}
[\Gamma(h \rightarrow Z\gamma)]_{HTM}
& = & \frac{\alpha G_F^2  m_W^2 m_{h}^3}
{64\pi^4} \left ( 1-\frac{m_Z^2}{m_h^2} \right )^3  \bigg|\frac{1}{c_W}  \sum_{f} 2 N^f_c Q_f (I_3^f-2Q_f s_W^2) g_{h ff} 
A^h_{1/2}
(\tau^h_f, \tau^Z_f)  \nonumber \\
 &+&\frac{(I_3^E-2Q_E s_W^2)(2Q_E)}{c_W} \bigg[\frac{\mu_{E_1}g_{hff}}{M_{E_1}} A_{1/2}(\tau^h_{E_1}, \tau^Z_{E_1}) +\frac{\mu_{E_2}g_{hff}}{M_{E_2}} A^h_{1/2}(\tau^h_{E_2}, \tau^Z_{E_2}) \bigg]  \nonumber \\
&+& c_W g_{h WW} A^h_1 (\tau^h_W, \tau^Z_W)- 2s_W \tilde{g}_{h H^\pm\,H^\mp}g_{ZH^\pm H^\mp}
A^h_0(\tau^h_{H^{\pm}}, \tau^Z_{H^\pm}) \nonumber \\
&-&
 4s_W \tilde{g}_{h H^{\pm\pm}H^{\mp\mp}} {g}_{Z H^{\pm\pm}H^{\mp\mp}}
A^h_0(\tau^h_{H^{\pm\pm}}, \tau^Z_{H^{\pm \pm}})        \bigg|^2 \, ,
\end{eqnarray}
where $\tau^h_i=4m_i^2/m_h^2,$ $\tau^Z_i=4m_i^2/m_Z^2$ (with $i=f(\equiv t),E_1, E_2, W,H^\pm,H^{\pm\pm}$), and the loop-factors are given by 
\begin{eqnarray}
A^h_0(\tau_h,\tau_Z) &=& I_1(\tau^h,\tau^Z),\nonumber \\
A^h_{1/2}(\tau^h,\tau^Z) &=& I_1(\tau^h,\tau^Z)-I_2(\tau^h,\tau^Z),\\
A^h_1(\tau^h,\tau^Z) &=& 4(3-\tan^2\theta_W)I_2(\tau^h,\tau^Z)+\left[(1+2\tau^{h\,-1})\tan^2\theta_W-(5+2\tau^{h\,-1})\right]I_1(\tau^h,\tau^Z).\nonumber
\end{eqnarray} 
The functions $I_1$ and $I_2$ are given by 
\begin{eqnarray}
I_1(\tau^h,\tau^Z) &=&  \frac{\tau^h \tau^Z}{2\left(\tau^h-\tau^Z\right)}
+\frac{\tau^{h \,2} \tau^{Z\,2}}{2\left(\tau^h-\tau^Z\right)^2}
\left[f\left(\tau^{h\,-1}\right)-f\left(\tau^{Z\,-1}\right)\right]\nonumber \\
&+&\frac{\tau^{h\,2}\tau^Z}{\left(\tau^h-\tau^Z\right)^2}
\left[g\left(\tau^{h\,-1}\right)-g\left(\tau^{Z\,-1}\right)\right],\nonumber\\
I_2(\tau^h,\tau^Z) &=& -\frac{\tau^h\tau^Z}{2(\tau^h-\tau^Z)}\left[f\left(\tau^{h\,-1}\right)-f\left(\tau^{Z\,-1}\right)\right],
\end{eqnarray}
where the function $f(\tau)$ is defined in Eq.~(\ref{eq:ftau}), and the function $g(\tau)$ is defined as
\begin{eqnarray}
g(\tau)=\left\{
\begin{array}{ll}  \displaystyle
\sqrt{\tau^{-1}-1}\sin^{-1}\left(\sqrt{\tau}\right), & (\tau< 1) \\
\displaystyle 
\frac{1}{2}\sqrt{1-\tau^{-1}}\left[ \log\left(\frac{1+\sqrt{1-\tau^{-1}}}
{1-\sqrt{1-\tau^{-1}}}\right)-i\pi\right], \hspace{0.5cm} & (\tau\geq 1) \, .
\end{array} \right. 
\label{eq:gtau} 
\end{eqnarray}
In Eq. (\ref{eq:hzg}) we list, in order,  the ordinary leptons, vector leptons, $W$ boson, singly-charged Higgs, and doubly-charged Higgs contribution. The scalar couplings $g_{hf\bar{f}}$ and $g_{hW^+W^-}$ are given in Eq.~(\ref{littleh1tt}), and the scalar trilinear couplings 
$\tilde{g}_{h H^{\pm}H^{\mp}}$ and $\tilde{g}_{h H^{\pm\pm}H^{\mp\mp}}$ are given in Eq.~(\ref{eq:redgcallittlehHp}). The 
remaining couplings in Eq.~(\ref{eq:hzg}) are given by 
\begin{eqnarray}
g_{Z H^{+} H^{-}} = - \tan \theta_W\, , \quad  g_{Z H^{++} H^{--}} = 2\cot 2\theta_W \, .
\end{eqnarray}
We proceed to perform a similar analysis as in Sec. \ref{subsec:gamgam}. We show first  the variation of the branching ratio $h \to Z \gamma$ with the mass $m^\prime_E=m_E^{\prime\prime}$, for various values of the doubly-charged Higgs masses, for the case of no mixing in the neutral sector, $\sin \alpha =0$ (shown in Fig. \ref{fig:hgZmE}, left-hand panel), and as a function of the mixing angle $\sin \alpha$ for $m^\prime_E=100$ GeV, and $m^\prime_E=200$ GeV, in the middle and right  panels of Fig. \ref {fig:hgZmE}, respectively. We have chosen the same parameter values as in Fig. \ref{fig:hgg}, for comparison.  It is clear that the branching ratio into $Z \gamma$ is fairly independent of both $m^\prime_E$ and $m_{H^{\pm \pm}}$, and always just below the SM expectations. Note that the severe suppression  seen in $h\to \gamma \gamma$ for $\sin \alpha=0$ (Fig. \ref{fig:hgg}, left side panel) does not occur here, and the results of the left side of Fig. \ref{fig:hgZmE} are consistent with the data at LHC.

But the variation with the mixing angle $\alpha$ is pronounced, and the branching ratio can reach almost twice its SM value for $\sin \alpha \sim 0.8$. However, correlated with our predictions from Sec. \ref{subsec:gamgam}  and LHC measurements for $R_{h \to \gamma \gamma}$, the parameter space corresponding to an enhanced $h \to Z \gamma$, for both $m_E^\prime=100$ GeV and 200 GeV, for  doubly-charged Higgs mass $m_{H^{\pm \pm}}=150$ GeV is ruled out. For all other values considered, the value for $R_{h \to Z \gamma}$ is close to, or below the SM expectations. This is general prediction of the model.

For a large range of parameter space, the decay $h \to Z \gamma$ can be suppressed significantly with respect to the SM. We plot decay rates for $h \to Z  \gamma $ as a function of $\sin \alpha$ for different values of doubly charged Higgs boson mass, considering $m_E^\prime =m_E^{\prime \prime}=100$ GeV and 200 GeV  in Fig.  \ref{fig:hgZmE}, middle and right panel respectively. Again, considerations for relative branching ratios are affected by the fact that the total width of Higgs boson in the HTM is not the same as in the SM. The widths are the same as those in the SM for $h$ for $\sin \alpha=0$, while for $\sin \alpha \ne 0$ we take into account the relative widths factors, Eq.(\ref{eq:width}).

In Tables \ref{tab:YZ1/6_ME_ML} and \ref{tab:YZ1/6_h_E} we present the explicit ranges of $R_{h \to Z \gamma}$ for varying $M_E, M_L$ and for a range of $h^\prime_E$ parameters. We choose a fixed value for $h^\prime_E=0.8$ in Table \ref{tab:YZ1/6_ME_ML}, as the preferred choice from other analyses \cite{Ishiwata:2011hr,Joglekar:2012vc}, and for comparison with Table \ref{tab:Y1/6_ME_ML}. Comparison of Tables \ref{tab:Y1/6_ME_ML} and \ref{tab:YZ1/6_ME_ML} shows that the decay $h \to Z \gamma$ is far more stable against variations in masses and values for $\sin \alpha$ than $h \to \gamma \gamma$, making it a less sensitive indicator for the presence of vector lepton states.

In Table \ref{tab:YZ1/6_h_E} we also allow variations in the Yukawa coupling $h^\prime_E\in (0-3)$. As before this amounts to allowing both explicit and contributions by electroweak symmetry breaking, $m_E^\prime,\, m_E^{\prime \prime}$, to vector lepton masses.
 Comparison of the Tables \ref{tab:YZ1/6_ME_ML} and \ref{tab:YZ1/6_h_E} indicates that the results are not very sensitive to variations in the Yukawa coupling $h^\prime_E$, or the vector lepton mass parameters $M_E, M_L$. However, the relative branching ratios are very sensitive to values of $\sin \alpha$. While the branching ratio into $Z \gamma$ is almost always suppressed with respect to its SM value, there is a small region of the parameter space, light $H^{\pm \pm}$ and  $\sin \alpha \simeq 0.7 -0.9$ where enhancement is possible; but as discussed before, this region is ruled out by constraints from $h \to \gamma \gamma$ measurement, Table \ref{tab:Y1/6_h_E}. Note that for $\sin \alpha=0$ the branching ratio is as before, independent of the mass of $H^{\pm \pm}$ and about the same as in the SM. 
\begin{table}[htbp]
\caption{\label{tab:YZ1/6_ME_ML}\sl\small Range of the ratio $R_{h \to Z \gamma}$, as defined in the text, for doubly-charged Higgs mass (in columns) and neutral Higgs mixing angle $\sin \alpha$ (in rows),  for Dirac vector lepton masses in the range $M_E, M_L \in (100-500)$ {\rm GeV},  with  $h^\prime_E=0.8$.}
  \begin{center}
 \footnotesize{
 \begin{tabular*}{0.98\textwidth}{@{\extracolsep{\fill}} c|cccccc}
 \hline\hline
  $R_{Z \gamma}$ &$m_{H^{\pm \pm}}=150$ GeV &$200$ GeV
 &$250$ GeV &$300$ GeV &$500$ GeV &$600$ GeV\\
  \hline\hline
  $\sin \alpha=0$ &$0.96-1$ &$0.96-1$ &$0.96-1$ &$0.96-1$ &$0.96-1$ &$0.96-1$  \\
  $\sin \alpha=0.1$ &$0.48-0.51$ &$0.68-0.72$ &$0.78-0.82$ &$0.83-0.87$  &$0.91-0.94$ &$0.92-0.96$ \\
  $\sin \alpha=0.2$ &$0.16-0.18$ &$0.44-0.47$  &$0.6-0.63$ & $0.69-0.73$  &$0.83-0.87$ &$ 0.85-0.89$  \\
  $\sin \alpha=0.3$ &$0.01-0.015$ &$0.24-0.26$  &$0.43-0.45$ &$0.55-0.58$  &$0.74-0.78$ &$0.78-0.81$   \\
  $\sin \alpha=0.4$ &$0.03-0.04$ &$0.09-0.10$  &$0.27-0.3$ &$0.41-0.43$  &$0.63-0.66$ &$0.68-0.71$ \\
  $\sin \alpha=0.5$ &$0.25-0.26$ &$0.01-0.02$  &$0.14-0.58$ &$0.27-0.29$  &$0.52-0.54$ &$0.56-0.59$   \\
  $\sin \alpha=0.6$ &$0.64-0.66$  &$0.005-0.006$  &$0.05-0.06$ &$0.15-0.17$ &$0.39-0.41$ & $0.43-0.46$ \\
  $\sin \alpha=0.7$ &$1.18-1.21$ &$0.07-0.08$  &$0.005-0.007$ &$0.06-0.07$  &$0.25-0.27$ &$0.3-0.32$    \\
  $\sin \alpha=0.8$ &$1.76-1.78$  &$0.21-0.22$  &$0.008-0.01$ &$0.009-0.011$ &$0.13-0.14$ &$0.17-0.18$  \\
  $\sin \alpha=0.9$  &$1.98-1.99$  &$0.35-0.36$ &$0.06$ &$0.004-0.005$ &$0.03-0.04$ &$0.05-0.06$ \\
    \hline
    \hline
   \end{tabular*}
   }
\end{center}
 \end{table}


%
\begin{table}[htbp]
\caption{\label{tab:YZ1/6_h_E}\sl\small Same as in Table \ref{tab:YZ1/6_ME_ML}, but also allowing  $ h^\prime_E \in (0-3)$.}
  \begin{center}
 \footnotesize{
 \begin{tabular*}{0.99\textwidth}{@{\extracolsep{\fill}} c|cccccc}
 \hline\hline
  $R_{Z \gamma}$ &$m_{H^{++}}=150$ GeV &$200$ GeV
 &$250$ GeV &$300$ GeV &$500$ GeV &$600$ GeV\\
  \hline\hline
  $\sin \alpha=0$ &$0.94-1.04$ &$0.94-1.04$ &$0.94-1.04$ &$0.94-1.04$ &$0.94-1.04$ &$0.94-1.04$  \\
  $\sin \alpha=0.1$ &$0.47-0.54$ &$0.66-0.74$ &$0.76-0.84$ &$0.8-0.9$  &$0.92-0.98$ &$0.9-1$ \\
  $\sin \alpha=0.2$ &$0.16-0.2$ &$0.43-0.49$  &$0.6-0.7$ & $0.68-0.74$  &$0.82-0.9$ &$ 0.8-0.9$  \\
  $\sin \alpha=0.3$ &$0.01-0.018$ &$0.23-0.27$  &$0.42-0.47$ &$0.54-0.6$  &$0.74-0.8$ &$0.76-0.84$   \\
  $\sin \alpha=0.4$ &$0.03-0.05$ &$0.09-0.12$  &$0.27-0.31$ &$0.4-0.5$  &$0.63-0.68$ &$0.67-0.73$ \\
  $\sin \alpha=0.5$ &$0.23-0.27$ &$0.01-0.02$  &$0.14-0.17$ &$0.27-0.3$  &$0.51-0.56$ &$0.56-0.61$   \\
  $\sin \alpha=0.6$ &$0.6-0.7$  &$0.001-0.01$  &$0.05-0.01$ &$0.15-0.18$ &$0.38-0.42$ & $0.43-0.47$ \\
  $\sin \alpha=0.7$ &$1.16-1.22$ &$0.07-0.08$  &$0.004-0.008$ &$0.06-0.08$  &$0.26-0.28$ &$0.3-0.33$    \\
  $\sin \alpha=0.8$ &$1.73-1.79$  &$0.2-0.23$  &$0.006-0.01$ &$0.008-0.01$ &$0.13-0.15$ &$0.17-0.18$  \\
  $\sin \alpha=0.9$  &$1.95-2$  &$0.34-0.36$ &$0.06-0.07$ &$0.004-0.005$ &$0.03-0.38$ &$0.05-0.06$ \\
    \hline
    \hline
   \end{tabular*}
   }
\end{center}
 \end{table}


\section{Production and Decays of the Doubly-Charged Higgs Bosons}
\label{sec:doublycharged}
The discovery of the doubly-charged Higgs bosons would be one of the most striking signals of physics beyond SM, and  a clear signature for the Higgs Triplet Model. The decay modes of $H^{\pm \pm}$ depend on the value of the VEV of the neutral triplet Higgs component, $v_\Delta$. When $v_\Delta \le 0.1$ MeV, the dominant decay mode of $H^{\pm \pm}$ is into lepton pairs. If $v_\Delta  \gg 0.1$ MeV, the main decay modes of $H^{\pm \pm}$ are into $W^{\pm (\star)} W^{\pm (\star)}$\footnote{ For the present analysis, the mass of the doubly-charged Higgs boson will be such that decays into on-shell $W^\pm$ pairs are kinematically allowed.}, and into  $H^{\pm} W^{\pm (\star)}$, if kinematically allowed. 
Searches for $H^{\pm \pm}$  were performed at Large Electron Positron
Collider (LEP) \cite{OPAL}, the Hadron Electron Ring Accelerator
(HERA) \cite{H1} and the Tevatron \cite{D0}.
The most up-to-date bounds have been more recently
derived by ATLAS and CMS collaborations at the LHC. Assuming a Drell-Yan-like pair production, these collaborations have looked for long-lived doubly-charged states, and after analyzing
5 fb$^{-1}$ of LHC collisions at a center-of-mass energy 
$\sqrt{s}=7$ TeV and 18.8 fb$^{-1}$ of collisions at $\sqrt{s}$ = 8 TeV,
they constrained the masses to lie above 685 GeV  \cite{Chatrchyan:2013oca}.
 The assumption is that the doubly-charged Higgs bosons decay 100\%
 into a pair of leptons with the same electric charge
through Majorana-type interactions \cite{Chatrchyan:2012ya}, thus neglecting possible decays into  $W$-boson pairs, shown to alter the pattern of $H^{\pm \pm}$ branching fractions \cite{Chiang:2012dk}. In this work, we allow both decays into $W^\pm W^\pm $ bosons, and also include  the effects of decays into vector leptons, which, if light enough, would modify the  decays of the doubly-charged Higgs bosons further. We take $v_\Delta=1$ GeV throughout our considerations\footnote{This value of $v_\Delta$ is small enough to satisfy electroweak precision conditions, but large enough to allow decay into gauge boson and charged Higgs \cite{Arbabifar:2012bd,Akeroyd:2007zv}. }.

The main production mode for $H^{\pm \pm}$ is the pair production $pp \to \gamma^*, Z^* \to H^{\pm \pm} H^{\mp \mp}$ and the associated production $pp \to W^{\pm *} \to H^{\pm \pm }H^{\mp}$. The production cross sections for both the  vector boson fusion $qQ \to q^\prime Q^\prime H^{\pm \pm}$, and for weak boson associated production $qQ \to W^{\pm\,*} \to H^{\pm \pm} W^\mp$, are proportional to $v_\Delta^2$ and much less significant for $v_\Delta \ll v_\Phi$.

At hadron colliders the partonic cross section for the leading order (LO) production cross section for doubly charged Higgs boson pair is
\begin{equation}
{\hat \sigma}_{LO}( q \bar q \to H^{\pm \pm } H^{\mp \mp})=\frac{\pi \alpha^2}{9Q^2}\beta^3 \left [4 e_q^2 +\frac{2e_q v_q v_{H^{\pm\pm}}(1-M_Z^2/Q^2)+(v_q^2+a_q^2)v_{H^{\pm \pm}}^2}{(1-M_Z^2/Q^2)^2+M_Z^2 \Gamma_Z^2/Q^4}\right],
\end{equation}
where we defined
\begin{eqnarray}
v_q&=&\frac{2I_{3q}-4e_q\sin^2 \theta_W}{\sin 2 \theta_W}, \qquad a_q=\frac{2I_{3q}}{\sin 2 \theta_W}, \qquad v_{H^{\pm \pm}}=\frac{2I_{3H^{\pm \pm}}-4\sin^2 \theta_W}{\sin 2 \theta_W},  \nonumber
\end{eqnarray}
with $I_{3i}$ the third component of the isospin for particle $i$,  $Q^2=\hat s$ the square of the partonic center of mass energy, $\displaystyle \beta =\sqrt{1-4m_{H^{\pm \pm}}^2/Q^2}$, and $\alpha$ the QED coupling constant evaluated at scale $Q$. The hadronic cross section is obtained by convolution with the partonic density functions of the proton
\begin{equation}
\sigma_{LO}( pp \to H^{\pm \pm } H^{\mp \mp})=\int_{\tau_0}^1 d\tau \sum_q\frac{d {\cal L}^{q \bar q}}{d \tau}{\hat \sigma}_{LO}(Q^2=\tau s),
\end{equation}
where ${\cal L}^{q \bar q}$ is the parton luminosity and  $\tau_0=4m_{H^{\pm\pm}}^2/s$ ($s$ is the total energy squared at the LHC). The cross section for pair-production, including NLO corrections, has been evaluated in \cite{Muhlleitner:2003me}.

Depending on mass parameters in the model, the doubly-charged Higgs boson can decay into lepton pairs, including vector leptons, $W^\pm$ pairs, or $H^\pm W^\pm$ states.  In the Higgs Triplet Model, the decay rate for $H^{\pm \pm}$ into leptons is 
\begin{eqnarray}
\Gamma (H^{\pm \pm} \to l_i^\pm l_j^\pm) =S_{ij} |h_{ij}|^2 \frac{m_{H^{\pm \pm}}}{4 \pi}\left ( 1-\frac{m_i^2}{m_{H^{\pm \pm}}^2}-\frac{m_j^2}{m_{H^{\pm \pm}}^2} \right ) \left [ \lambda \left(\frac{m_i^2}{m_{H^{\pm \pm}}^2}, \frac{m_j^2}{m_{H^{\pm \pm}}^2} \right) \right]^2, 
\end{eqnarray}
where $m_i$ is the mass of the $i-$th lepton ($i=e, \mu$ or $\tau$) and $S_{ij}=1, (1/2)$ for $i \ne j$, ($i=j$).
Similarly the decay rate of  $H^{\pm \pm}$ into fourth generation vector leptons is, if kinematically allowed 
\small
\begin{eqnarray}
\Gamma (H^{\pm \pm} \to E_i^\pm E_j^\pm) =S_{ij}\left[  |h^\prime_{E_iE_j}|^2+|h_{E_iE_j}^{\prime \prime}|^2 \right] \frac{m_{H^{\pm \pm}}}{4 \pi}\left ( 1-\frac{m_{E_i}^2}{m_{H^{\pm \pm}}^2}-\frac{m_{E_j}^2}{m_{H^{\pm \pm}}^2} \right ) \left [ \lambda \left(\frac{m_{E_i}^2}{m_{H^{\pm \pm}}^2}, \frac{m_{E_j}^2}{m_{H^{\pm \pm}}^2} \right) \right]^2, \nonumber \\
\end{eqnarray}
\normalsize
with $M_{E_i}$ the mass eigenvalue from Eq. (\ref{eq:eigenvalues}). In addition, we use the decay rates of $H^{\pm \pm}$ into $W^\pm W^\pm$ and $W^\pm H^\pm$:
\begin{eqnarray}
\Gamma(H^{\pm \pm} \to W^\pm W^\pm)&=&\frac{g^4v_\Delta^2 m_{H^{\pm \pm}}^3}{64  \pi m_W^4} \left ( 1- \frac{4 m_W^2}{m_{H^{\pm \pm}}^2} + \frac{12 m_W^4}{m_{H^{\pm \pm}}^4} \right) \beta \left (\frac{m_W^2}{m_{H^{\pm \pm}}^2} \right) \\
\Gamma(H^{\pm \pm} \to W^\pm H^\pm)&=&\frac{g^2 m_{H^{\pm \pm}}^3}{16  \pi m_W^2} \cos^2\beta_{\pm} \left [ \lambda \left( \frac{m_W^2}{m_{H^{\pm \pm}}^2},  \frac{m_{H^\pm}^2}{m_{H^{\pm \pm}}^2}\right) \right]^{3/2}, 
\end{eqnarray}
where $\cos \beta_\pm \simeq 1$ is the mixing angle in the singly-charged Higgs sector and
\begin{equation}
\beta(x) = \sqrt{1-4x}, \qquad \lambda (x,y)= 1+x^2+y^2-2xy-2x-2y.
\end{equation}

\begin{figure}[t]
\center
\vskip -0.6in 
\begin{center}
$
\begin{array}{cc}
\vspace*{-0.6in}
\hspace*{-1.2cm}
\includegraphics[width=3.5in,height=3.5in]{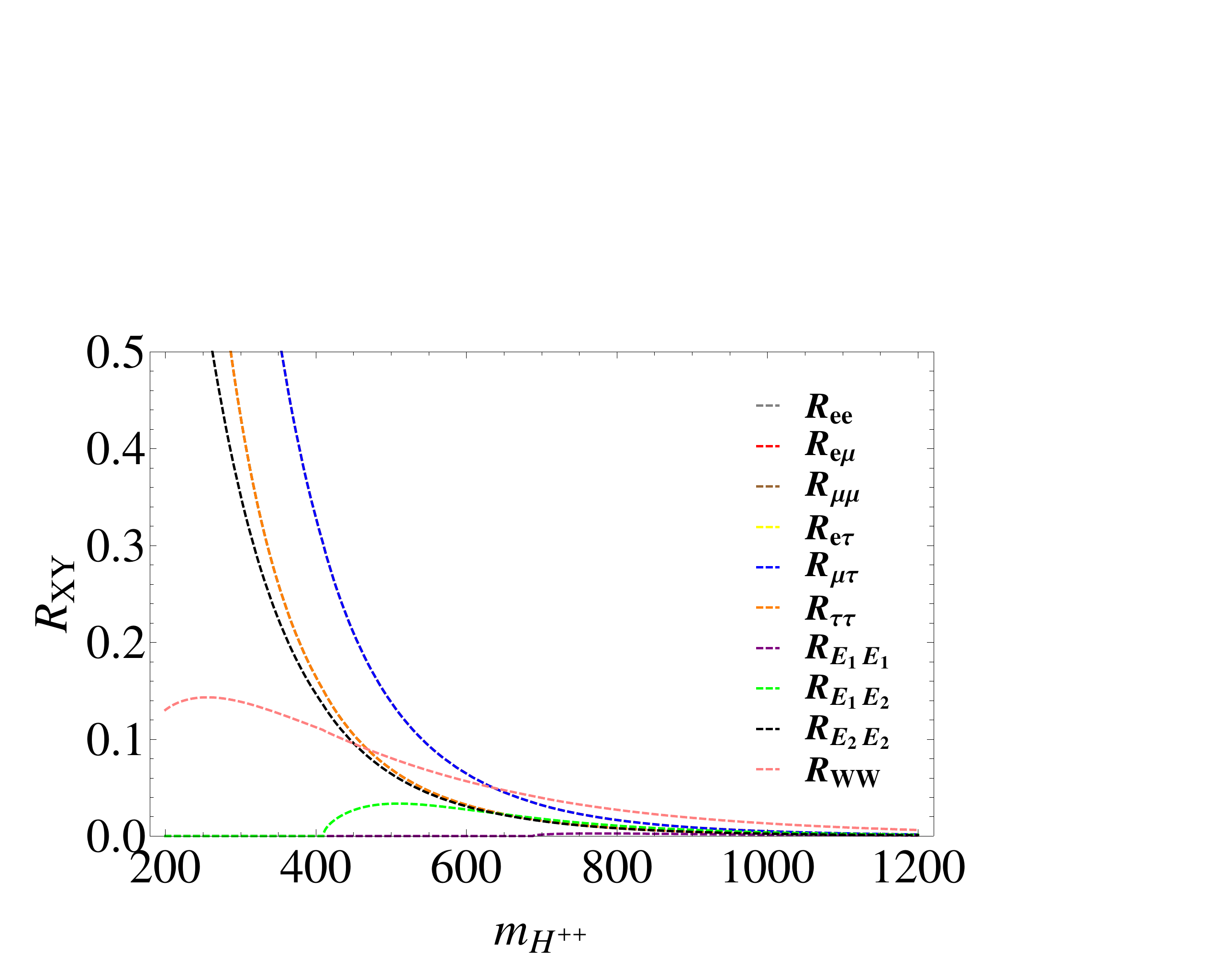}
&\hspace*{-0.8cm}
	\includegraphics[width=3.5in,height=3.5in]{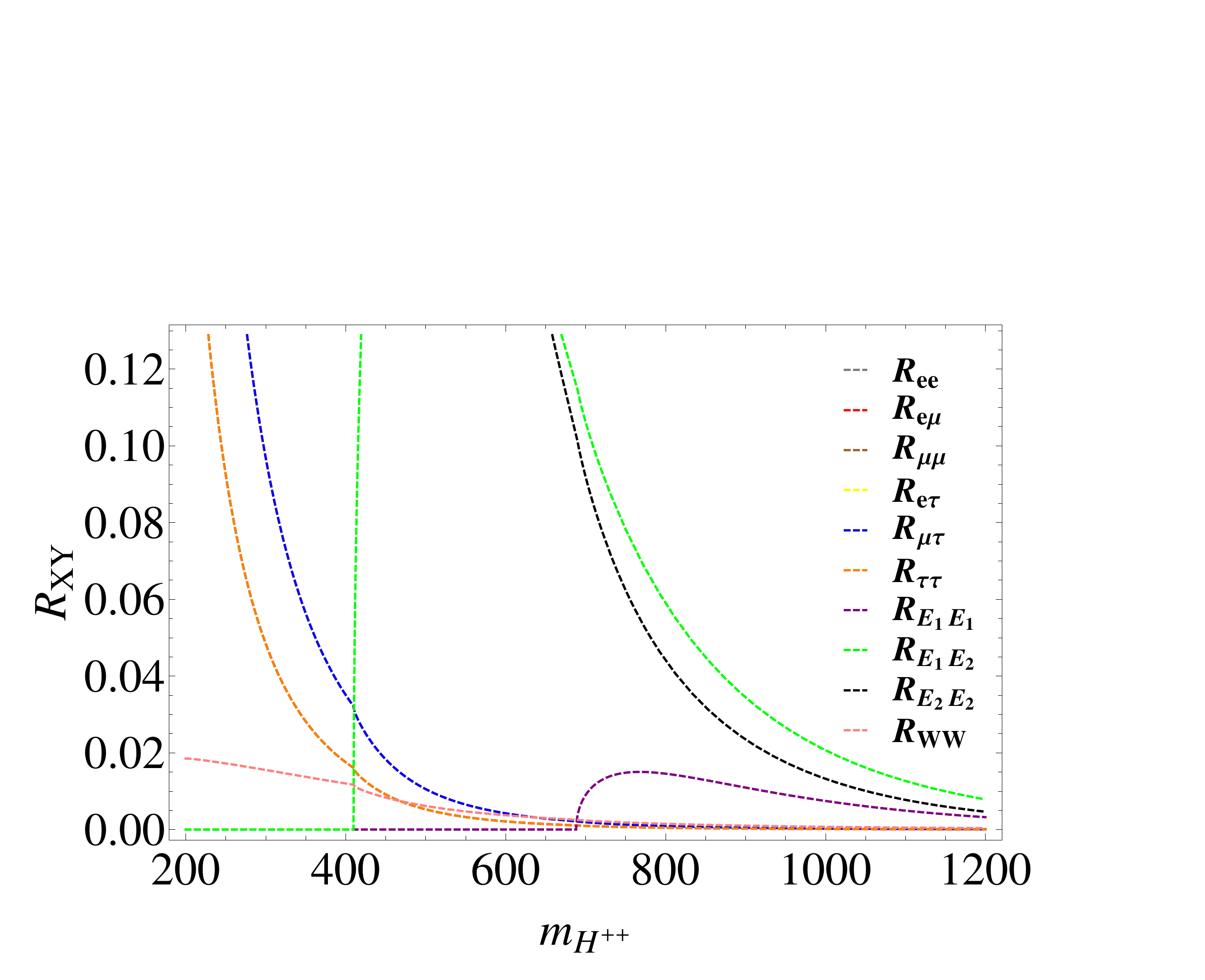}\\	 
\hspace*{-1.2cm}
	\includegraphics[width=3.5in, height=3.5in]{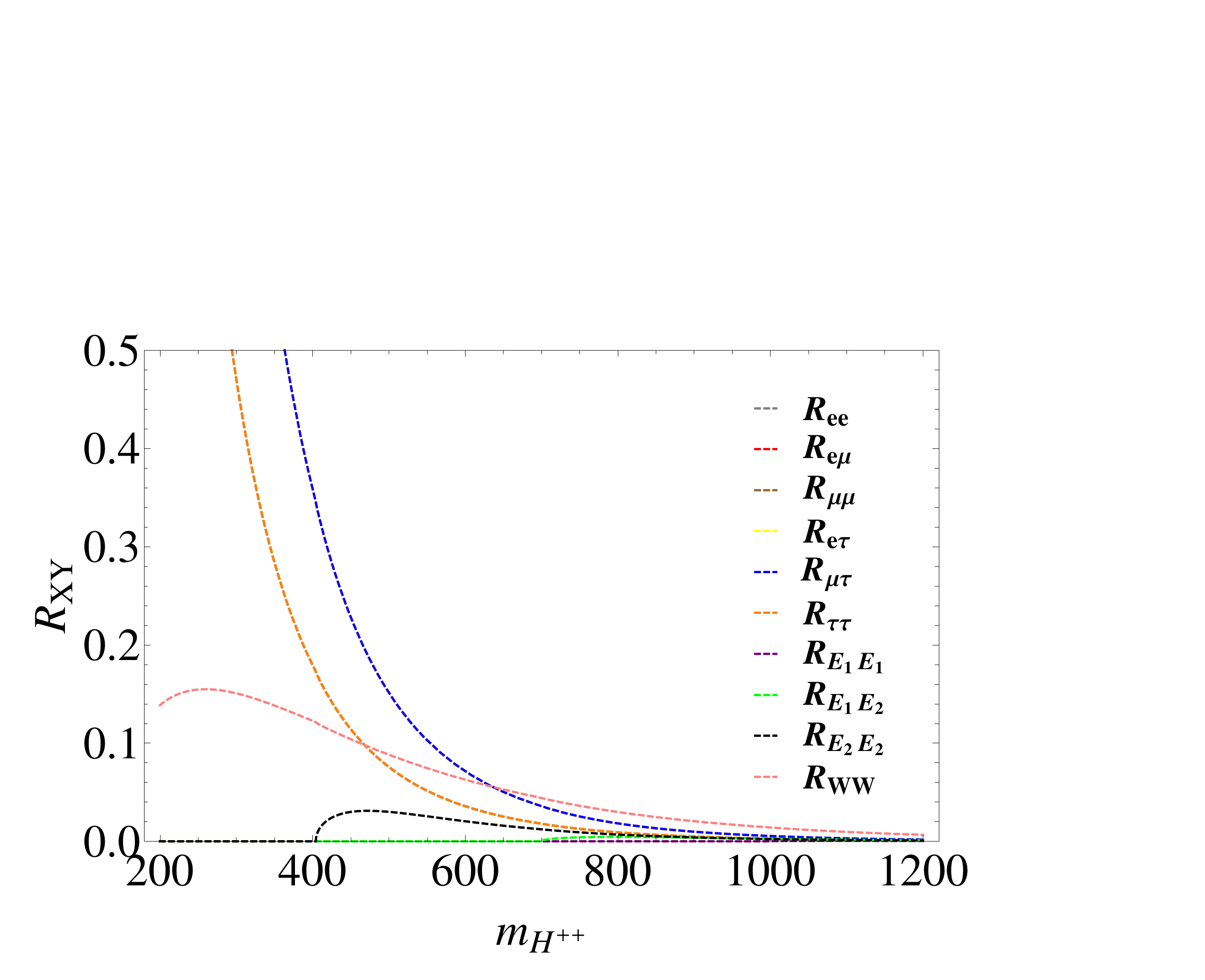}
&\hspace*{-0.8cm}
	\includegraphics[width=3.5in,height=3.5in]{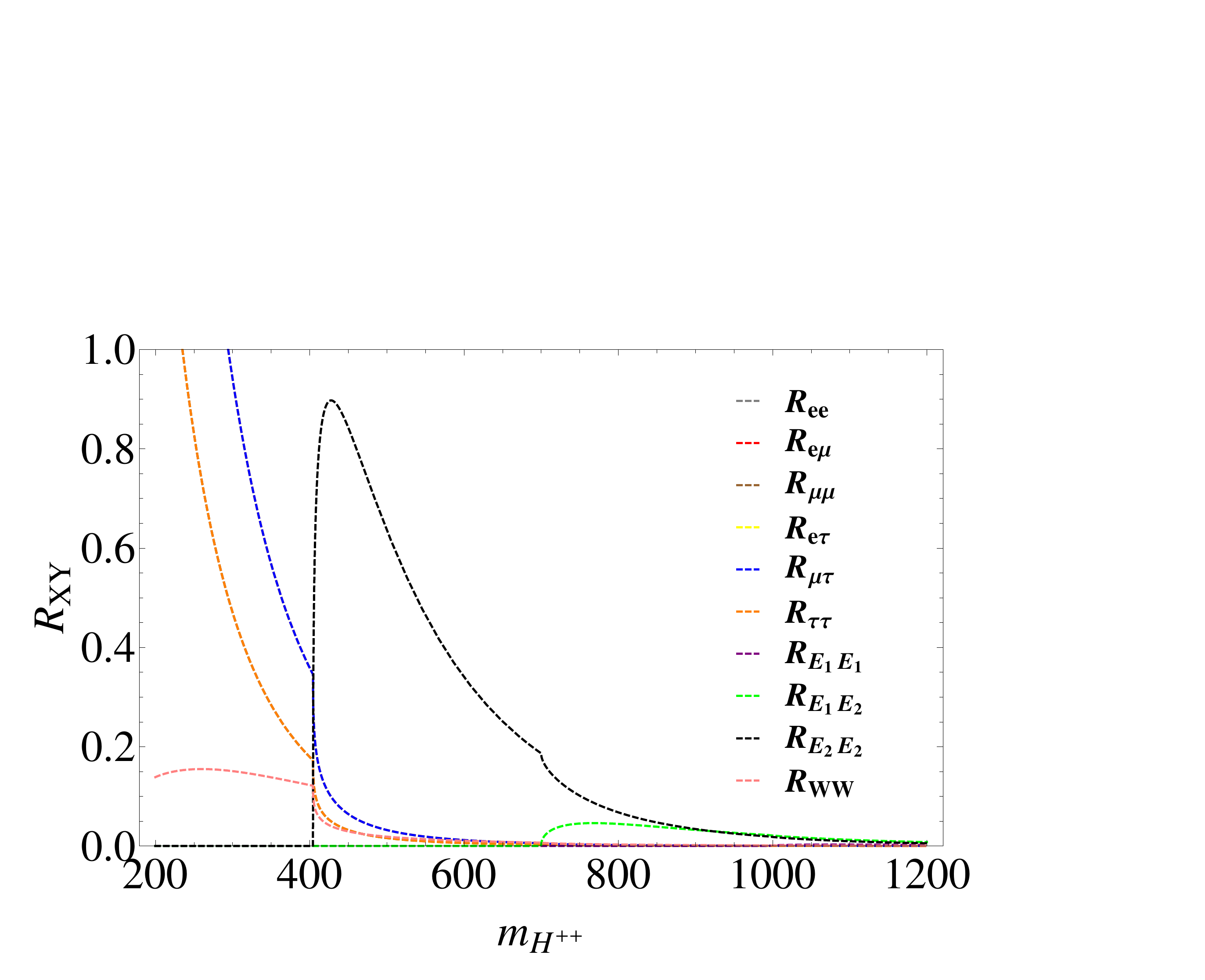}
        \end{array}$
        \end{center}
\caption{$R_{XY}=\sigma (pp\to H^{\pm \pm} H^{\mp \mp}) \times BR(H^{\pm \pm} \to XY)$ as a function of doubly charged Higgs boson mass satisfying Condition 1: (top left) Case A and $h^\prime_{EE}=h_{EE}^{\prime \prime}=0.01$; (top right) Case A and $h^\prime_{EE}=h_{EE}^{\prime \prime}=0.1$; (bottom left)  Case B and $h^\prime_{EE}=h_{EE}^{\prime \prime}=0.01$; (bottom right) Case B and $h^\prime_{EE}=h_{EE}^{\prime \prime}=0.1$. We take $h_{ij}=0.01$. } 
\label{fig:condition1}
\end{figure}

We investigate the branching ratios of $H^{\pm \pm}$ in two distinct parameter regions:
\begin{itemize}
\item Condition 1: when $H^{\pm \pm} \to W^\pm H^\pm$ is not kinematically allowed. Then $H^{\pm \pm}$ decays into leptons and $W^\pm$ pairs only:
\small
\begin{eqnarray}
BR( X_i^\pm X_j^\pm)=\frac{ \Gamma (H^{\pm \pm} \to X_i^\pm X_j^\pm) } {\Gamma_1(H^{\pm \pm})}, \qquad {\rm where} \quad X_i=l_i^\pm, E_i^\pm, W^\pm,
\end{eqnarray}
with the total width for Condition 1 $$\Gamma_1(H^{\pm \pm}) = \Gamma (H^{\pm \pm} \to l_i^\pm l_j^\pm) +\Gamma (H^{\pm \pm} \to E_i^\pm E_j^\pm) + \Gamma (H^{\pm \pm} \to W^\pm W^\pm). $$ 
\normalsize
\item Condition 2: when $H^{\pm \pm} \to W^\pm H^\pm$ is  kinematically allowed.  Then $H^{\pm \pm}$ is able to decay into charged Higgs and gauge bosons as well:
\small
\begin{eqnarray}
BR( X_i^\pm X_j^\pm)&=&\frac{ \Gamma (H^{\pm \pm} \to X_i^\pm X_j^\pm) } {\Gamma_2(H^{\pm \pm})} \qquad {\rm where} \quad X_i=l_i^\pm, E_i^\pm, W^\pm\,, \qquad {\rm and}\nonumber\\
BR( W^\pm H^\pm)&=&\frac{ \Gamma (H^{\pm \pm} \to H^\pm W^\pm)}{\Gamma_2(H^{\pm \pm})},
\end{eqnarray}
with the total decay width for Condition 2 $$\Gamma_2(H^{\pm \pm}) = \Gamma (H^{\pm \pm} \to l_i^\pm l_j^\pm) +\Gamma (H^{\pm \pm} \to E_i^\pm E_j^\pm) + \Gamma (H^{\pm \pm} \to W^\pm W^\pm)+\Gamma (H^{\pm \pm} \to H^\pm W^\pm). $$ 
 \normalsize
\end{itemize}
We investigate the decay patterns of $H^{\pm \pm}$ and present plots of the production cross section times the branching fractions under various conditions. To cover a wide range of parameter space, we distinguish  two cases for each condition, depending on the vector lepton masses.  We set the Yukawa coupling of the vector leptons with the doublet Higgs bosons to be  ${ h}_E^\prime= {h}_E^{\prime \prime}=0.8$ for both cases. 
\begin{itemize}
\item Case A corresponds to very light vector leptons:   $M_E=M_L=205$ GeV. For this case we obtain for the eigenvalues, $M_{E_1}=344.2$ GeV, $M_{E_2}=65.8$ GeV, the latter of which is close to the allowed minimum. 
\item Case B corresponds to intermediate mass vector leptons  $M_E=400$ GeV, $M_L=300$ GeV. For this case we obtain for the eigenvalues, $M_{E_1}=498$ GeV, $M_{E_2}=202$ GeV.
\end{itemize}
In Fig. \ref{fig:condition1} we show the corresponding graphs for Condition 1 (when the decay $H^{\pm \pm} \to W^\pm H^\pm$ is not kinematically allowed), Case A on the top row and Case B on the bottom row. We plot  $R_{XY}=\sigma (pp\to H^{\pm \pm} H^{\mp \mp}) \times BR(H^{\pm \pm} \to XY)$ with $X, Y$ as specified in the attached panels, as functions of the doubly-charged masses. On the left side of the figure   the Yukawa couplings of vector lepton with the triplet Higgs boson  are $h^\prime_{EE}=h_{EE}^{\prime \prime}= 0.1$ and on the right side of the figure, $h^\prime_{EE}=h_{EE}^{\prime \prime}=0.01$. We set $h_{ij}=0.01$ throughout.

For small vector lepton coupling ${ h}^\prime_{EE}=0.01$, the decay into $\mu^\pm \tau^\pm$ is dominant at low $H^{\pm \pm}$ masses (this is because we assumed  the couplings $h_{ij}$ with ordinary leptons to be all equal; should we have chosen them diagonal, the branching ratio into $\tau^\pm \tau^\pm$ would dominate). At high $H^{\pm\pm}$ masses  the branching ratio into $W^\pm W^\pm$ dominates and can reach 40\%, branching ratios into ordinary leptons reaching at most 40\%, and those into vector leptons remaining at or below 20\% level. If the triplet Yukawa coupling to vector  leptons is allowed to increase to $h^\prime_{EE}=h_{EE}^{\prime \prime}=0.1$, the decay  into the two lightest vector leptons $E_2$ becomes dominant, and can reach 80-90\% when kinematically allowed (in the $m_{H^{\pm\pm}} \sim 200-400$ GeV region for $M_{E_2}=65.8$ GeV) and overwhelms the other decay modes, which are now below 5\%. The only difference between Case A and Case B in this figure are threshold effects. For Case B, $m_{H^{\pm\pm}}>400$ GeV for decay into pairs of $E_2$ states, as $M_{E_2}=202$ GeV; otherwise the branching ratios are the same.

\begin{figure}[t]
\center
\vskip -0.6in 
\begin{center}
$
	\begin{array}{cc}
	\vspace*{-0.6in}
\hspace*{-1.2cm}
\includegraphics[width=3.5in,height=3.5in]{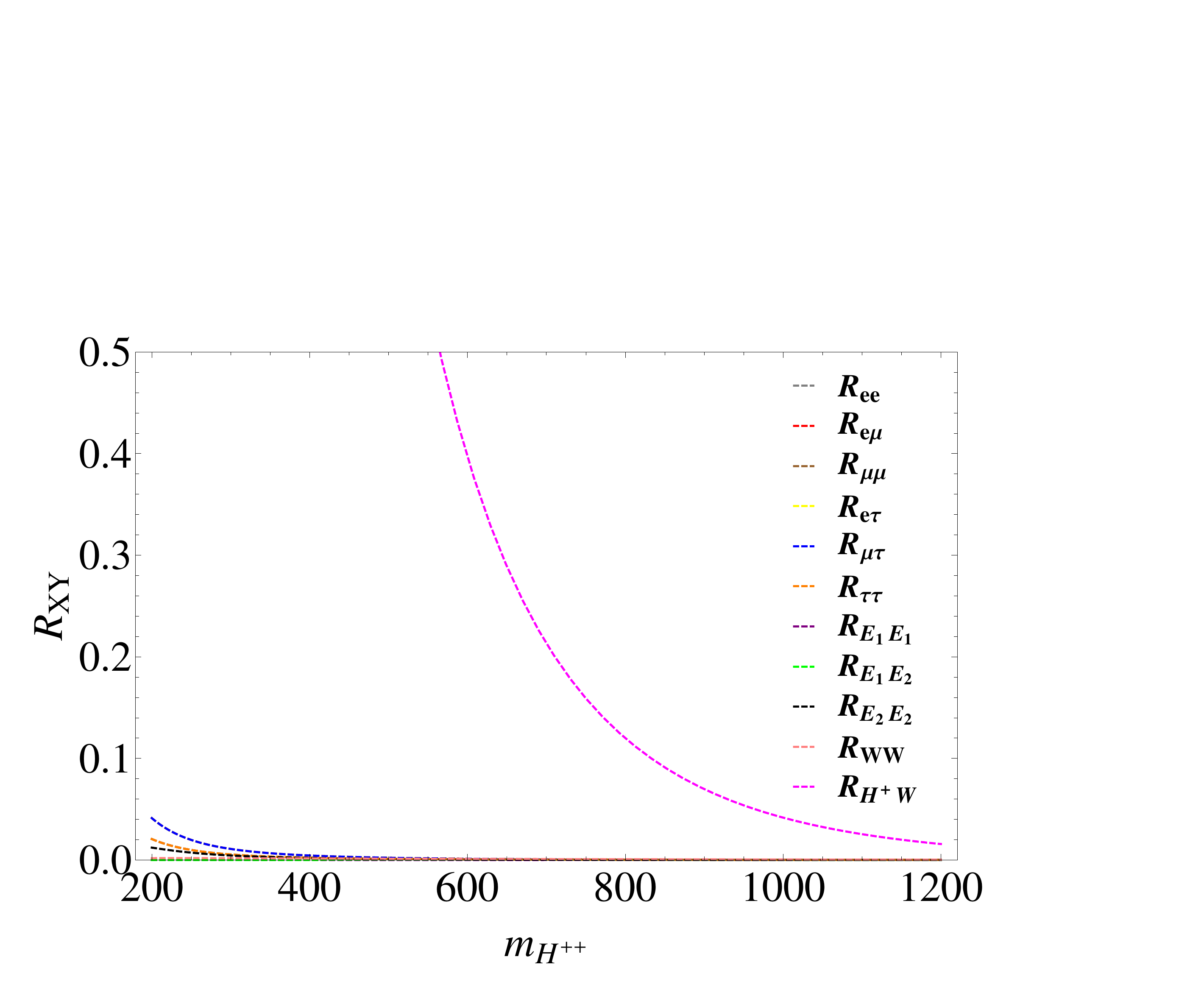}
&\hspace*{-0.6cm}
\includegraphics[width=3.5in,height=3.5in]{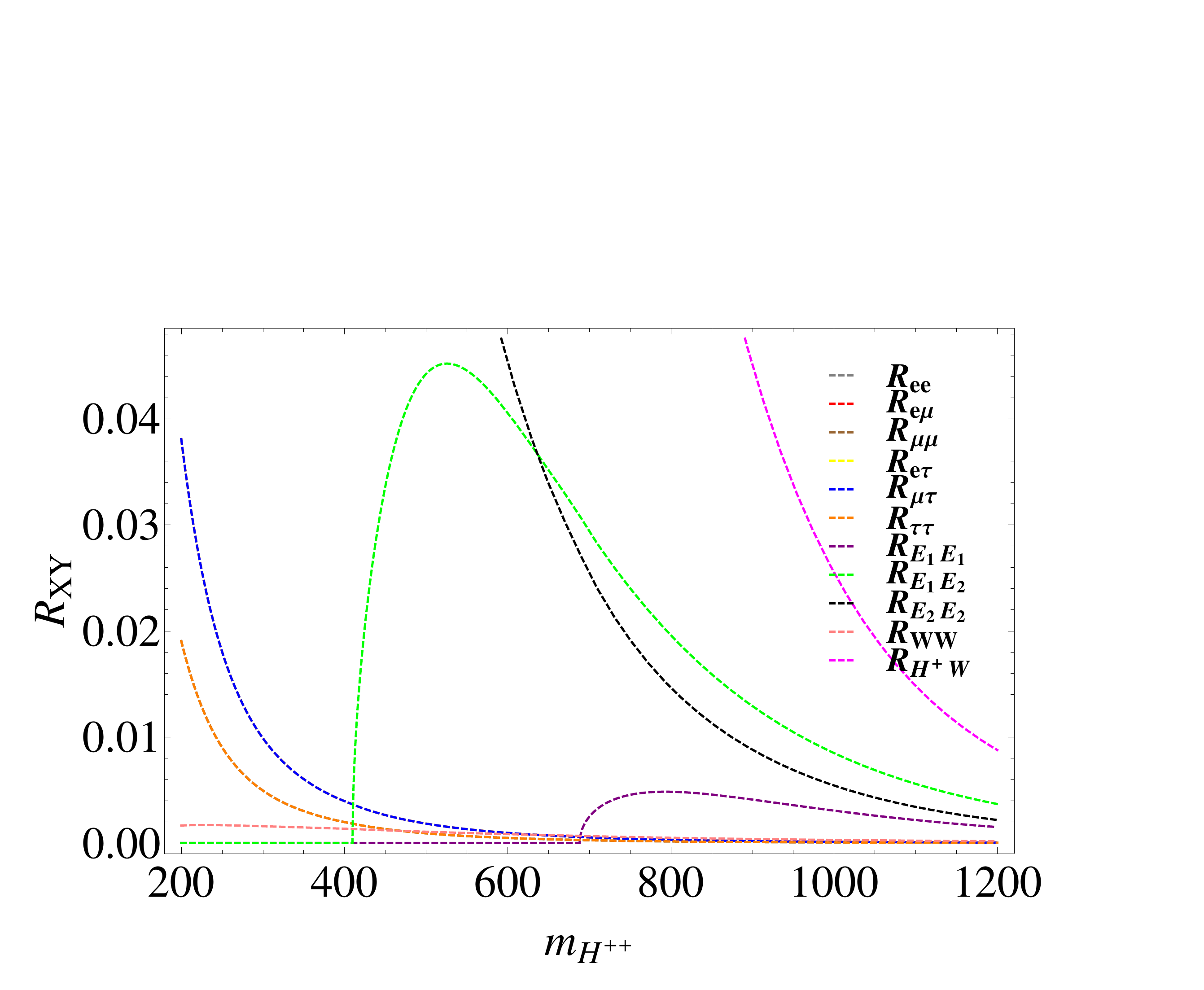}\\
\hspace*{-1.2cm}
	\includegraphics[width=3.5in, height=3.5in]{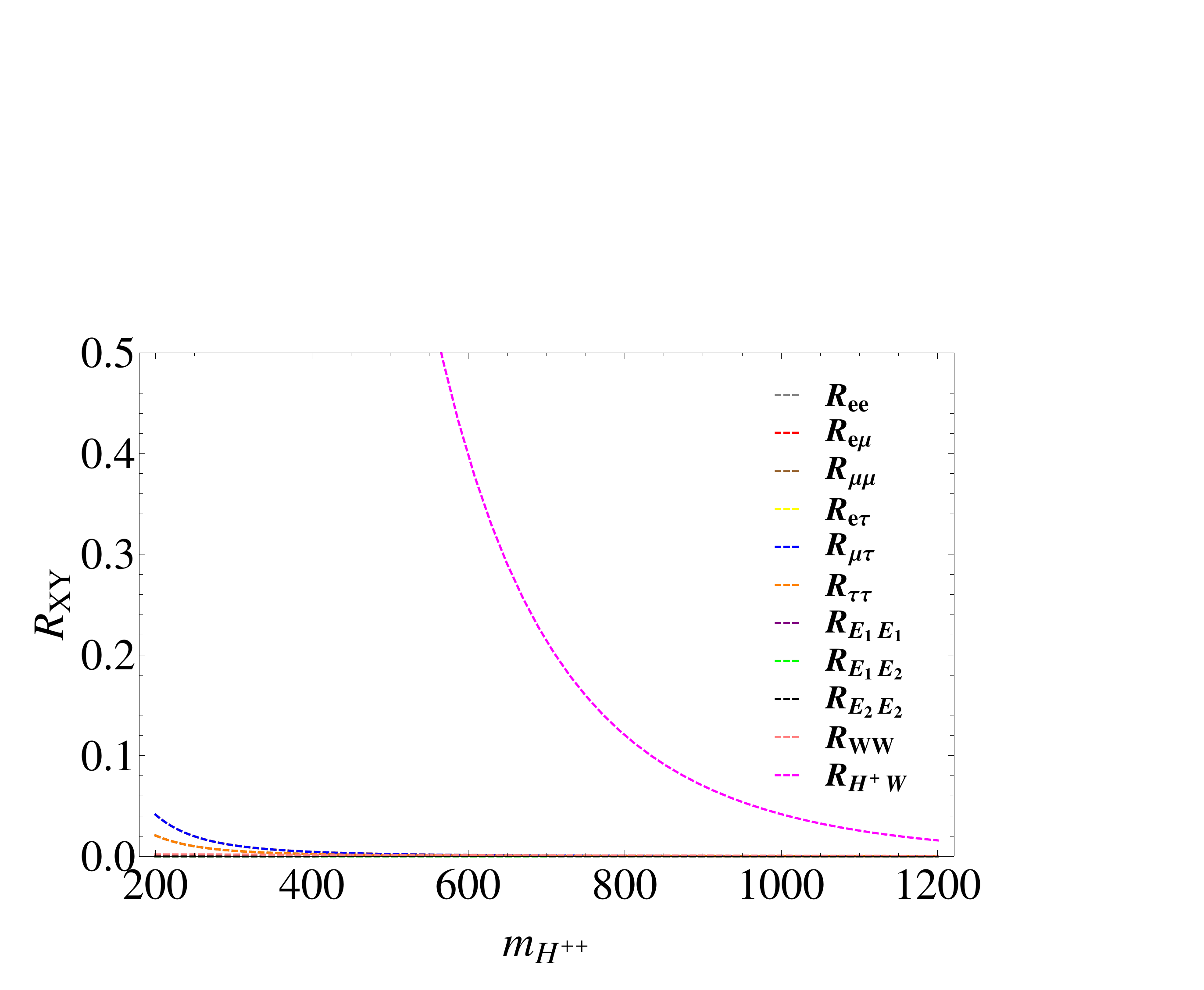}
&\hspace*{-0.6cm}
	\includegraphics[width=3.5in,height=3.5in]{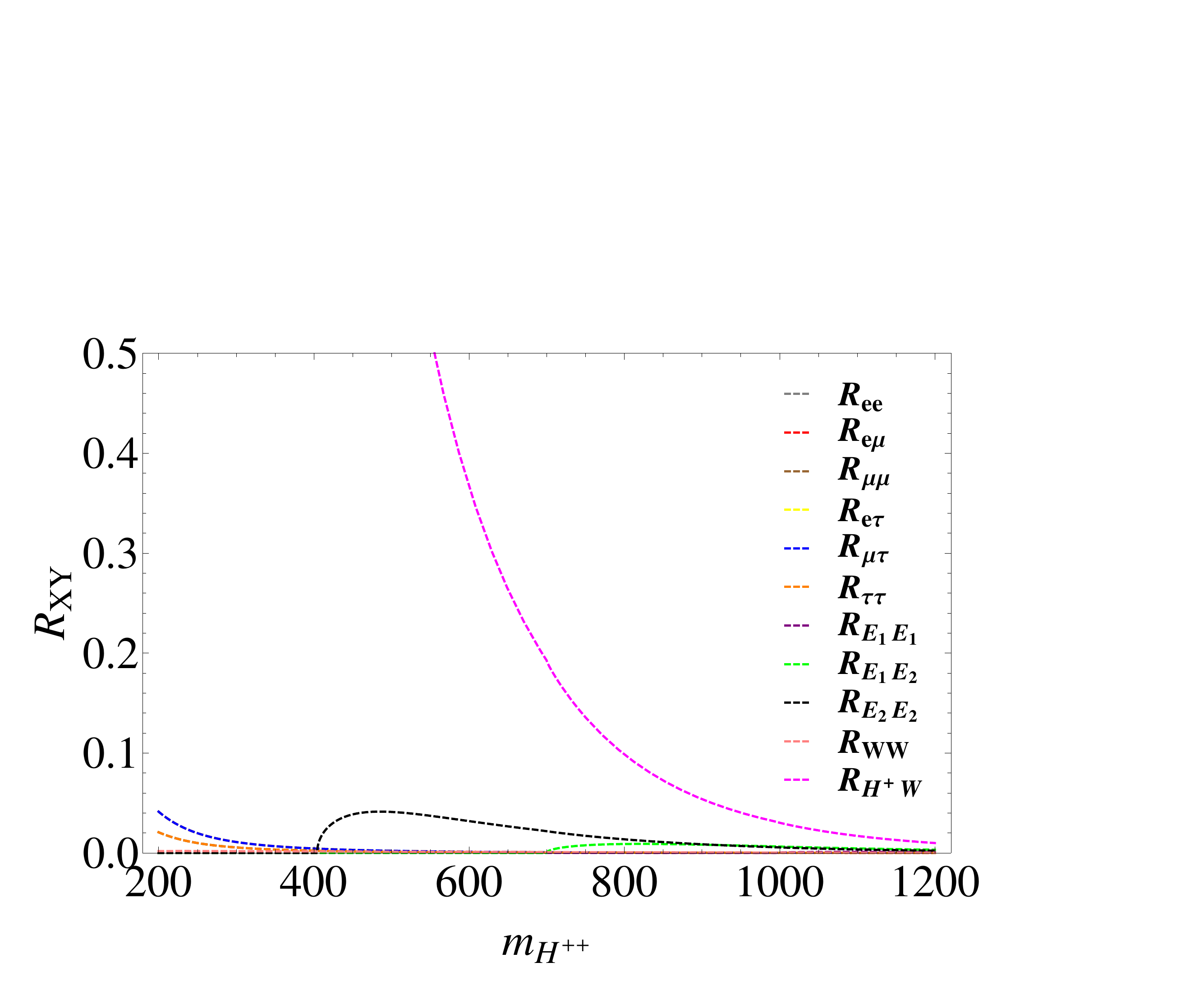}
        \end{array}$
        \end{center}
\caption{$R_{XY}=\sigma (pp\to H^{\pm \pm} H^{\mp \mp}) \times BR(H^{\pm \pm} \to XY)$ as a function of doubly charged Higgs boson mass satisfying Condition 2: (top left) Case A and $h^\prime_{EE}=h_{EE}^{\prime \prime}=0.01$ (top right); Case A and $h^\prime_{EE}=h_{EE}^{\prime \prime}=0.1$; (bottom left) Case B and $h^\prime_{EE}=h_{EE}^{\prime \prime}=0.01$; (bottom right) Case B and $h^\prime_{EE}=h_{EE}^{\prime \prime}=0.1$. We take throughout  $h_{ij}=0.01$ and $m_{H^\pm}=120$ GeV for Condition 2.} 
\label{fig:condition2}
\end{figure}

 In Fig. \ref{fig:condition2} we plot the same quantities for Condition 2  (when $H^{\pm \pm} \to W^\pm H^\pm$ is  kinematically allowed), also Case A on the top row and case B on the bottom row, $h^\prime_{EE}=h_{EE}^{\prime \prime}=0.01$ on the left hand side and $h^\prime_{EE}=h_{EE}^{\prime \prime}=0.1$ on the right hand side.  The decay pattern is very different here, and it is dominated by $H^{\pm \pm} \to W^\pm H^\pm$. 
 For  small vector lepton coupling $h^\prime_{EE}=h_{EE}^{\prime \prime}$ the branching ratios for decays into vector and ordinary lepton pairs, and $W^\pm$ pairs are very small, and reach at most 1\%.   Increasing the vector lepton coupling  to $h^\prime_{EE}=h_{EE}^{\prime \prime}=0.1$, the decay  into $H^\pm W^\pm$ still dominates throughout the parameter space where it is kinematically allowed and can reach a branching fraction of over 90\%, while the decay into vector leptons can have branching ratios of up to 25\%. Again, the decay rates into $W^\pm$ boson pairs and ordinary leptons are below 1\%, and the only difference between Case A and Case B are, as in Fig. \ref{fig:condition1}, threshold effects.

\clearpage

\section{Conclusion}
\label{sec:conclusion}
Despite no new signals of physics beyond the SM at the LHC, the SM cannot be the  complete theory of particle interactions. An extension of the SM by additional vector-like leptons is not ruled out experimentally, and has been shown to provide a dark matter candidate. In models beyond the SM, the vector leptons can alter  not only the phenomenology of the Higgs, but also of other additional particle representations predicted by the models. We provide an example within the Higgs Triplet Model, where previously we showed that, in the absence of triplet-doublet Higgs mixing in the neutral sector ($\sin \alpha=0$), there is no enhancement of the rate of decay of $h \to \gamma \gamma$ in this model with respect to the SM expectation.

Introducing vector leptons does not affect any of the tree-level decays of the neutral Higgs boson observed at the LHC, nor its production decay. However, loop decays into electroweak  particles, such as $h \to Z \gamma$ and $h \to \gamma \gamma$ would be affected. We show that, for the no-mixing scenario ($\sin \alpha=0$) the decays rates into $\gamma \gamma$ and $Z \gamma$ do not depend on the doubly charged Higgs mass, and thus without the additional vector leptons, these decays would be unchanged from the SM expectations. With vector leptons, modest enhancements or suppressions are possible, more so for $h \to \gamma \gamma$, where for large Yukawa couplings, the rate of decays could even double. Under the same circumstances, the decay width for $h \to Z \gamma$ remains practically unchanged from its SM value. The model thus presents a mechanism for enhancing one loop-decays and not the other, which seems to be consistent with the LHC data (so far). 

If $\sin \alpha \ne 0$, the effect of the doubly-charged Higgs bosons  is felt for both $ h \to \gamma \gamma$ and $ h \to Z \gamma$, most spectacularly so for very light $m_{H^{\pm \pm}}=150$ GeV, which is ruled out for $\sin \alpha\ne 0$. Parameter space regions where light doubly-charged Higgs masses  (200-250 GeV), {\it and} significant mixing in the neutral sector coexist, are disfavored. In general, there are many parameter combinations for which the decay $ h \to \gamma \gamma$ is enhanced, but few for an enhanced $ h\to Z \gamma$, and these regions are ruled out by the branching ratio for $h \to \gamma \gamma$. However, if the decay $ h \to \gamma \gamma$ is (modestly) enhanced, while $h \to Z \gamma$ is the same as the SM to 1$\sigma$, small mixing angles and light doubly charged Higgs bosons $m_{H^{\pm \pm}} \lsim 300$ GeV are preferred. The fact that there are no regions of the parameter space consistent with present measurements of the branching ratio for $h \to \gamma \gamma$ and an enhanced rate for $h \to Z \gamma$ is a feature of this model, valid over the whole explored range of the parameter space. Other than that, there are no definite correlations or anti-correlations between there two loop-dominated decays.

The intermediate mass doubly-charged Higgs boson can decay into light vector leptons, which would alter its decay profile significantly.  We explored this possibility and found that, if the singly charged Higgs mass is such that the decay $H^{\pm \pm} \to W^\pm H^\pm$ is not kinematically accessible, dominant branching ratios into vector leptons, if kinematically accessible, are expected for  triplet Yukawa couplings $h^\prime_{EE}=0.1$, whereas for small $h^\prime_{EE}=0.01$ branching ratios into ordinary leptons, vector leptons and $W^\pm$ pairs are comparable for $m_{H^{\pm \pm}} \ge 600$ GeV. If and where the decay $H^{\pm \pm} \to W^\pm H^\pm$ is kinematically accessible, its corresponding branching ratio is the largest, while the branching fraction into vector leptons could reach  20-25\% for $h^\prime_{EE}=0.1$.  Under both these circumstances, the decay patterns of the doubly charged Higgs bosons are significantly altered, raising the hope that they can be found at masses around 300-500 GeV.
\acknowledgments
The work of S. B. and M.F.  is supported in part by NSERC under grant number SAP105354. 



\begin{thebibliography}{99}

\bibitem{HiggsLHC}
  G.~Aad {\it et al.}  [ATLAS Collaboration],
  Phys.\ Lett.\ B {\bf 716} (2012) 1;
  S.~Chatrchyan {\it et al.}  [CMS Collaboration],
  Phys.\ Lett.\ B {\bf 716} (2012) 30.
  
  \bibitem{Aaltonen:2009nr}
  T.~Aaltonen {\it et al.}  [CDF Collaboration],
  Phys.\ Rev.\ Lett.\  {\bf 104}, 091801 (2010);
  A.~Lister  [CDF Collaboration],
  arXiv:0810.3349 [hep-ex].


  
\bibitem{Agashe:2004cp} 
  K.~Agashe, G.~Perez and A.~Soni,
  Phys.\ Rev.\ D {\bf 71}, 016002 (2005);
  K.~Agashe, G.~Perez and A.~Soni,
  Phys.\ Rev.\ D {\bf 75}, 015002 (2007);
  A.~Carmona and J.~Santiago,
  JHEP {\bf 1201}, 100 (2012);
  G.~-Y.~Huang, K.~Kong and S.~C.~Park,
  JHEP {\bf 1206}, 099 (2012);
  C.~Biggio, F.~Feruglio, I.~Masina and M.~Perez-Victoria,
  Nucl.\ Phys.\ B {\bf 677}, 451 (2004);
  D.~E.~Kaplan and T.~M.~P.~Tait,
  JHEP {\bf 0006} (2000) 020;
  H.~-C.~Cheng,
  Phys.\ Rev.\ D {\bf 60}, 075015 (1999).

  
  
\bibitem{Moroi:1991mg}
  T.~Moroi and Y.~Okada,
  Mod.\ Phys.\ Lett.\ A {\bf 7} (1992) 187;
  T.~Moroi and Y.~Okada,
  Phys.\ Lett.\ B {\bf 295} (1992) 73; 
  K.~S.~Babu, I.~Gogoladze, M.~U.~Rehman and Q.~Shafi,
  Phys.\ Rev.\ D {\bf 78} (2008) 055017; 
  S.~P.~Martin,
  Phys.\ Rev.\ D {\bf 81} (2010) 035004; 
  P.~W.~Graham, A.~Ismail, S.~Rajendran and P.~Saraswat,
  Phys.\ Rev.\ D {\bf 81} (2010) 055016; 
  S.~P.~Martin,
  Phys.\ Rev.\ D {\bf 82} (2010) 055019;
  J.~Kang, P.~Langacker and B.~D.~Nelson,
  Phys.\ Rev.\ D {\bf 77} (2008) 035003.
  
\bibitem{Dobrescu:1997nm}
  B.~A.~Dobrescu and C.~T.~Hill,
  Phys.\ Rev.\ Lett.\  {\bf 81} (1998) 2634; 
  R.~S.~Chivukula, B.~A.~Dobrescu, H.~Georgi and C.~T.~Hill,
  Phys.\ Rev.\ D {\bf 59} (1999) 075003;
  H.~Collins, A.~K.~Grant and H.~Georgi,
  Phys.\ Rev.\ D {\bf 61} (2000) 055002; 
  H.~-J.~He, C.~T.~Hill and T.~M.~P.~Tait,
  Phys.\ Rev.\ D {\bf 65} (2002) 055006;
  C.~T.~Hill and E.~H.~Simmons,
  Phys.\ Rept.\  {\bf 381} (2003) 235
   [Erratum-ibid.\  {\bf 390} (2004) 553];
  R.~Contino, L.~Da Rold and A.~Pomarol,
  Phys.\ Rev.\ D {\bf 75} (2007) 055014;
  Phys.\ Rev.\ D {\bf 79} (2009) 075003;
  K.~Kong, M.~McCaskey and G.~W.~Wilson,
  JHEP {\bf 1204} (2012) 079;
  A.~Carmona, M.~Chala and J.~Santiago,
  JHEP {\bf 1207} (2012) 049;
  M.~Gillioz, R.~Grober, C.~Grojean, M.~Muhlleitner and E.~Salvioni,
  JHEP {\bf 1210} (2012) 004.

\bibitem{ArkaniHamed:2002qy}
  N.~Arkani-Hamed, A.~G.~Cohen, E.~Katz and A.~E.~Nelson,
  JHEP {\bf 0207} (2002) 034;
  T.~Han, H.~E.~Logan, B.~McElrath and L.~-T.~Wang,
  Phys.\ Rev.\ D {\bf 67} (2003) 095004;
  M.~Perelstein, M.~E.~Peskin and A.~Pierce,
  Phys.\ Rev.\ D {\bf 69} (2004) 075002;
  M.~Schmaltz and D.~Tucker-Smith,
  Ann.\ Rev.\ Nucl.\ Part.\ Sci.\  {\bf 55} (2005) 229;
  M.~S.~Carena, J.~Hubisz, M.~Perelstein and P.~Verdier,
  Phys.\ Rev.\ D {\bf 75} (2007) 091701:
  S.~Matsumoto, T.~Moroi and K.~Tobe,
  Phys.\ Rev.\ D {\bf 78} (2008) 055018.
  
\bibitem{Davidson:1987tr}
  A.~Davidson and K.~C.~Wali,
  Phys.\ Rev.\ Lett.\  {\bf 60} (1988) 1813;
  K.~S.~Babu and R.~N.~Mohapatra,
  Phys.\ Rev.\ D {\bf 41} (1990) 1286;
  B.~Grinstein, M.~Redi and G.~Villadoro,
  JHEP {\bf 1011} (2010) 067;
  D.~Guadagnoli, R.~N.~Mohapatra and I.~Sung,
  JHEP {\bf 1104} (2011) 093.

  
\bibitem{Azatov:2009na} 
  A.~Azatov, M.~Toharia and L.~Zhu,
  Phys.\ Rev.\ D {\bf 80}, 035016 (2009).
  
  \bibitem{VFlim}
  ATLAS Collaboration ATLAS-CONF-2013-019; S. Chatrchyan et al. [CMS Collaboration], Phys.\ Lett.\ B {\bf 718}, 348 (2012). 
  
\bibitem{delAguila:2008pw} 
  F.~del Aguila, J.~de Blas and M.~Perez-Victoria,
  Phys.\ Rev.\ D {\bf 78}, 013010 (2008).
  
\bibitem{Joglekar:2012vc} 
  A.~Joglekar, P.~Schwaller and C.~E.~M.~Wagner,
  JHEP {\bf 1212}, 064 (2012).
 

 
\bibitem{Kearney:2012zi} 
  J.~Kearney, A.~Pierce and N.~Weiner,
  Phys.\ Rev.\ D {\bf 86}, 113005 (2012);
  H.~Davoudiasl, H.~-S.~Lee and W.~J.~Marciano,
  Phys.\ Rev.\ D {\bf 86}, 095009 (2012);
  K.~J.~Bae, T.~H.~Jung and H.~D.~Kim,
  Phys.\ Rev.\ D {\bf 87}, 015014 (2013);
  M.~B.~Voloshin,
  Phys.\ Rev.\ D {\bf 86}, 093016 (2012);
  D.~McKeen, M.~Pospelov and A.~Ritz,
  Phys.\ Rev.\ D {\bf 86}, 113004 (2012);
  H.~M.~Lee, M.~Park and W.~-I.~Park,
  JHEP {\bf 1212}, 037 (2012);
  C.~Arina, R.~N.~Mohapatra and N.~Sahu,
  arXiv:1211.0435 [hep-ph];
  B.~Batell, S.~Jung and H.~M.~Lee,
  JHEP {\bf 1301}, 135 (2013);
  H.~Davoudiasl, I.~Lewis and E.~Ponton,
  arXiv:1211.3449 [hep-ph].;
  J.~Fan and M.~Reece,
  arXiv:1301.2597 [hep-ph];
  A.~Carmona and F.~Goertz,
  arXiv:1301.5856 [hep-ph];
  C.~Cheung, S.~D.~McDermott and K.~M.~Zurek,
  arXiv:1302.0314 [hep-ph];
  W.~-Z.~Feng and P.~Nath,
  arXiv:1303.0289 [hep-ph];
  C.~Englert and M.~McCullough,
  arXiv:1303.1526 [hep-ph].
    
\bibitem{Ishiwata:2011hr} 
  K.~Ishiwata and M.~B.~Wise,
  Phys.\ Rev.\ D {\bf 84}, 055025 (2011).
  
\bibitem{Garg:2013rba} 
  S.~K.~Garg and C.~S.~Kim,
  arXiv:1305.4712 [hep-ph].


  
  

   
\bibitem{Joglekar:2013zya} 
  A.~Joglekar, P.~Schwaller and C.~E.~M.~Wagner,
  arXiv:1303.2969 [hep-ph].
  
\bibitem{Giudice:2008uua} 
  G.~F.~Giudice and O.~Lebedev,
  Phys.\ Lett.\ B {\bf 665}, 79 (2008); 
  E.~Del Nobile, R.~Franceschini, D.~Pappadopulo and A.~Strumia,
  Nucl.\ Phys.\ B {\bf 826}, 217 (2010); 
  S.~P.~Martin,
  Phys.\ Rev.\ D {\bf 81}, 035004 (2010).
  
    
\bibitem{Arbabifar:2012bd} 
  F.~Arbabifar, S.~Bahrami and M.~Frank,
  Phys.\ Rev.\ D {\bf 87}, 015020 (2013).
  
  \bibitem{ATLASnew}
  [CMS Collaboration], CMS-PAS-HIG-13-001 and CMS-PAS-HIG-13-005; 
G. Aad et al. [ATLAS Collaboration], [arXiv:1307.1427 [hep-ex]] and ATLAS-CONF 2013-029.
  
\bibitem{Chatrchyan:2012ya} 
  S.~Chatrchyan {\it et al.}  [CMS Collaboration],
  Eur.\ Phys.\ J.\ C {\bf 72}, 2189 (2012);
  G.~Aad {\it et al.}  [ATLAS Collaboration],
  Eur.\ Phys.\ J.\ C {\bf 72}, 2244 (2012).
  
  
  
  \bibitem{Akeroyd:2007zv}
A.~G. Akeroyd, Mayumi Aoki, and Hiroaki Sugiyama.
 Phys. Rev. D {\bf 77}, 075010  (2008);
A.~G. Akeroyd, Mayumi Aoki, and Hiroaki Sugiyama.
 Phys. Rev.  D {\bf 79}, 113010 (2009);
Takeshi Fukuyama, Hiroaki Sugiyama, and Koji Tsumura.
 JHEP {\bf 03}, 044 (2010);
S.~T. Petcov, H.~Sugiyama, and Y.~Takanishi.
 Phys. Rev. D {\bf 80}, 015005 (2009);
Takeshi Fukuyama, Hiroaki Sugiyama, and Koji Tsumura.
Phys. Rev. D {\bf 82}, 036004 (2010);
A.G. Akeroyd, Cheng-Wei Chiang, and Naveen Gaur,
JHEP {\bf 1011}, 005 (2010);
  A.~Arhrib, R.~Benbrik, M.~Chabab, G.~Moultaka and L.~Rahili,
  JHEP {\bf 1204}, 136 (2012);
  A.~Arhrib, R.~Benbrik, M.~Chabab, G.~Moultaka and L.~Rahili,
  [arXiv:1202.6621 [hep-ph]];
  A.~Arhrib, R.~Benbrik, M.~Chabab, G.~Moultaka, M.~C.~Peyranere, L.~Rahili and J.~Ramadan,
  Phys.\ Rev.\ D {\bf 84}, 095005 (2011);
E.~J.~Chun, H.~M.~Lee and P.~Sharma,
JHEP {\bf 1211}, 106 (2012);
A. Melfo, M. Nemev\v sek, F. Nesti, G. Senjanovi\'c and Y. Zhang,
Phys. Rev. D 85,  055018 (2012);
  M.~Aoki, S.~Kanemura and K.~Yagyu,
  Phys.\ Rev.\ D {\bf 85}, 055007 (2012);
  S.~Kanemura and K.~Yagyu,
  Phys.\ Rev.\ D {\bf 85}, 115009 (2012);
  M.~Aoki, S.~Kanemura, M.~Kikuchi and K.~Yagyu,
  [arXiv:1211.6029 [hep-ph]].
  

 
\bibitem{Ishiwata:2013gma} 
  K.~Ishiwata and M.~B.~Wise,
  arXiv:1307.1112 [hep-ph].
  
\bibitem{Dermisek:2013gta} 
  R.~Dermisek and A.~Raval,
  arXiv:1305.3522 [hep-ph]. 

\bibitem{Chatrchyan:2013vaa} 
  S.~Chatrchyan {\it et al.}  [CMS Collaboration],
  arXiv:1307.5515 [hep-ex];
  [ATLAS Collaboration], ATLAS-CONF-2013-009.
  
\bibitem{Chen:2013vi} 
  C.~-S.~Chen, C.~-Q.~Geng, D.~Huang and L.~-H.~Tsai,
  Phys.\  Rev.\  D87, {\bf 075019} (2013);
  C.~-S.~Chen, C.~-Q.~Geng, D.~Huang and L.~-H.~Tsai,
  Phys.\ Lett.\ B {\bf 723}, 156 (2013);
  P.~S.~Bhupal Dev, D.~K.~Ghosh, N.~Okada and I.~Saha,
  JHEP {\bf 1303}, 150 (2013)
  [Erratum-ibid.\  {\bf 1305}, 049 (2013)].
  

  
  
  
  
   \bibitem{OPAL}
   G.~Abbiendi {\it et al.}  [OPAL Collaboration],
  Phys.\ Lett.\  B {\bf 526} (2002) 221;
G.~Abbiendi {\it et al.}  [OPAL Collaboration],
  Phys.\ Lett.\  B {\bf 577} (2003) 93;
  P.~Achard {\it et al.}  [L3 Collaboration],
  Phys.\ Lett.\  B {\bf 576} (2003) 18;
J.~Abdallah {\it et al.}  [DELPHI Collaboration],
  Phys.\ Lett.\  B {\bf 552} (2003) 127.


\bibitem{H1}
  J.~Sztuk-Dambietz [H1 and ZEUS Collaboration],
  J.\ Phys.\ Conf.\ Ser.\  {\bf 110} (2008) 072042;
  A.~Aktas {\it et al.}  [H1 Collaboration],
  Phys.\ Lett.\  B {\bf 638} (2006) 432;
  
\bibitem{D0}
V.~M.~Abazov {\it et al.}  [D0 Collaboration],
  Phys.\ Rev.\ Lett.\  {\bf 101} (2008) 071803;
  V.~M.~Abazov {\it et al.}  [D0 Collaboration],
  Phys.\ Rev.\ Lett.\  {\bf 108} (2012) 021801;
  D.~E.~Acosta {\it et al.}  [CDF Collaboration],
  Phys.\ Rev.\ Lett.\  {\bf 93} (2004) 221802;
 T.~Aaltonen {\it et al.}  [The CDF Collaboration],
  Phys.\ Rev.\ Lett.\  {\bf 101} (2008) 121801.

\bibitem{Chatrchyan:2013oca} 
  S.~Chatrchyan {\it et al.}  [ CMS Collaboration],
  arXiv:1305.0491 [hep-ex];
  G.~Aad {\it et al.}  [ ATLAS Collaboration],
  Phys.\ Lett.\ B 722 (2013) 305;
  S.~Chatrchyan {\it et al.}  [CMS Collaboration],
  Eur.\ Phys.\ J.\ C {\bf 72} (2012) 2189;
  G.~Aad {\it et al.}  [ATLAS Collaboration],
  Eur.\ Phys.\ J.\ C {\bf 72} (2012) 2244.

\bibitem{Chiang:2012dk} 
  C.~-W.~Chiang, T.~Nomura and K.~Tsumura,
  Phys.\ Rev.\ D {\bf 85}, 095023 (2012);
  S.~Kanemura, K.~Yagyu and H.~Yokoya,
  arXiv:1305.2383 [hep-ph].
  
\bibitem{Muhlleitner:2003me} 
  M.~Muhlleitner and M.~Spira,
  Phys.\ Rev.\ D {\bf 68}, 117701 (2003).

\end{thebibliography}
\end{document}